\documentclass{aa}
\def\gr{$\gamma$-ray}
\usepackage{natbib} 
\usepackage{graphicx} 
\usepackage{color}
\begin{document}

\title{Catalog of very-high-energy emitting active galactic nuclei at high Galactic latitudes}
\author{Andrii Neronov$^{1,2}$, Dmitri Semikoz$^{1}$}

\institute{
Universit\'e Paris Cit\'e, CNRS, Astroparticule et Cosmologie, 75006 Paris, France
\and
Laboratory of Astrophysics, \'Ecole Polytechnique F\'ed\'erale de Lausanne, 1015 Lausanne, Switzerland
}

\authorrunning{Neronov and Semikoz}
\titlerunning{VHE AGN catalog}

 \abstract
     {Large number of Active Galactic Nuclei (AGN) producing Very-High-Energy (VHE) gamma-rays (energies above 100~GeV) has been revealed using observations with Imaging Atmospheric Cherenkov Telescopes (IACTs). However, our knowledge of the VHE emitting AGN population is limited because the IACTs have not yet performed a systematic extragalactic sky survey.}{ We use long exposure of Fermi Large Area Telescope (LAT) to perform a survey of VHE emitting AGN at Galactic latitudes $|b|>10^\circ$.}{We consider clustering of gamma-ray events with energies $E>100$ GeV around positions of AGN from the 4-th LAT source catalog to select sources detected in the VHE range. }{The VHE AGN catalog produced in this way contains overall 175 sources detected with high confidence and additional 100 sources detected at more than $3\sigma$ level. It is 90\% complete at the flux limit $1.3\times 10^{-12}$~erg/cm$^2$s. Less than half of the source sample (71) are previously reported VHE emitters, other sources are new detections in the VHE band. The majority of VHE AGN detectable at the survey flux limit are BL Lac type objects. We find their luminosity function to derive their spatial density $(6.5\pm 0.5)\times 10^{-7}$~Mpc$^{-3}$ and the characteristic luminosity scale $\simeq 10^{44}$~erg/s. Ten sources in the VHE AGN catalog are nearby radio galaxies and seven are flat spectrum radio quasars, while 20 sources are unclassified AGN. We also include in our catalog four unidentified sources that may or may not be VHE AGN. 63 source in the catalog are "extreme" blazars, with 41 of them being new VHE band detections. In spite of the fact that the VHE flux is heavily attenuated by the pair production in interactions with Extragalactic Background Light (EBL), the catalog includes 7 sources at redshift larger than 1. Some of these sources show peculiar hardening of the VHE band spectra that point either to errors in redshift determination, or to limitations of modeling of cosmological evolution of EBL. }{}

\maketitle

\section{Introduction}

Properties of the Very-High Energy (VHE, photon energies above 100~GeV) emitting extragalactic source population(s) are poorly known because of the limitations of the observational technique of Imaging Atmospheric Cherenkov Telescopes (IACT) used for observations in the VHE domain. The IACT systems have relatively narrow Field-of-View  (FoV) of several degrees across and mostly do not perform large sky surveys because of limited observation time available over the year and relatively long exposures needed for source detections. The best effort so far has been done by HESS collaboration that has produced an extragalactic sky survey \citep{2025A&A...695A.261H} by searching for serendipitous source detection in available HESS telescope pointings toward high Galactic latitude directions (Galactic latitudes above $|b|=10^\circ$). The extragalactic sky survey by HESS covers only 5.7\% of the sky, with no new source detections in addition to sources already known from dedicated pointed observations. Another limitation of the IACT sky surveying technique is the absence of a regular sampling of the sky. Because of this, it is currently impossibly to know the time-average flux of most of the known extragalactic sources that are typically variable on a range of time scales.

VHE \gr\ flux from distant extragalactic sources is absorbed on the way from the source to the Earth by the effect of pair produciton on Extragalactic Backgorund Light (EBL), a collective optical-infrared emission from all galaxies accumulated all over the history of the Universe \citep{2006Natur.440.1018A,2017A&A...603A..34F,2021MNRAS.507.5144S}. This phenomenon leads to the occurrence of "gamma-ray horizon": limited maximal distance of extragalactic sources observable with telescopes operating in the VHE band. The highest redshift source that has been detected up to now above 100~GeV is OP 313 at redshift $z=0.997$ \citep{2023ATel16381....1C}. Existence of the gamma-ray horizon potentially biases selection of sources for pointed observations with IACTs: high-redshift sources have lower chances to be detectable with IACTs and hence are not selected for such observations. 

The current state-of-art of the knowledge of the extragalactic VHE sky is summarized in the TeVCat\footnote{https://www.tevcat.org/}, a list of VHE detected sources compiled from scientific literature. TeVCat currently lists 106 VHE sources at high Galactic latitudes $|b|>10^\circ$. This list is not a catalog in a proper sense, and it cannot be used for the characterization of the population of extragalactic VHE sources, because entries of TeVCat are not chosen using well-defined selection criteria. 

In spite of much smaller collection area, compared to IACTs, space-based Fermi Large Area Telescope (LAT) \citep{2009ApJ...697.1071A} is also detecting VHE sources. First attempts of the all-sky searches of extragalactic VHE sources have been reported soon after the start of operations of Fermi/LAT \citep{2011A&A...529A..59N,2015A&A...575A..21N} using relatively short telescope exposure. These studies have shown the potential of unbiased sky survey approach for the study of VHE sources, by discovering a number of sources at high redshift and detecting sources that have not been reported by IACTs. The most recent Fermi/LAT source catalog \citep{Ballet:2023qzs} reports measurements of source fluxes in energy bands including the VHE energy range. Most of the Fermi catalog sources are not detected in the VHE range, because of the limited sensitivity of Fermi/LAT. 

Nevertheless, the Fermi/LAT sensitivity at 100~GeV in a more-than-ten year exposure is approximately equal to that of the new IACT facility in construction now, the Cherenkov Telescope Observatory (CTAO) \citep{2019scta.book.....C}. Prospects for AGN study with CTAO have been previously assessed via extrapolation of  LAT \gr\ spectra from low energies $E<100$ GeV to the VHE band \cite{CTAConsortium:2013tmf}. Long LAT exposure removes the need for extrapolation: most of the sources with flux above the CTAO sensitivity limit at 100~GeV should now be detected by Fermi/LAT.

In what follows we present the catalog of VHE emitting AGN detected by Fermi/LAT in the high Galactic latitude sky ($|b|>10^\circ$) in a sixteen-year-long exposure. We use this catalog to characterize the population of the VHE emitting AGN, the dominant source class in this part of the sky. 

\section{Fermi/LAT data analysis}

For our analysis we use Fermi/LAT publicly available data in the time range between September 2008 and March 2025 and energy range above 100~GeV. We select events from the P8R3\_SOURCEVETO\_V3 class that have the lowest residual cosmic ray contamination. Apart from the residual cosmic ray background, an additional background for detection of isolated VHE sources is provided by the Galactic diffuse emission. We consider the sky regions at Galactic latitude $|b|>10^\circ$ to suppress this diffuse emission background.

We use the selected dataset to find clusters of VHE events around sources from the 4th Fermi/LAT source catalog, 4FGL \citep{Ballet:2023qzs}. There are ${\cal N}_{100}=13251$ photons with energy above 100~GeV in the high Galactic latitude regions ($|b|>10^\circ$). The chance coincidence probability of one photon to fall within $\theta=0.1^\circ$ around a source (approximately the 68\% containment circle of the Point-Spread-Function, PSF) is $p_1={\cal N}_{100}\theta^2/(4(1-\cos(80^\circ)))\simeq 0.0122$.
The probability for a source to have three photons by chance
$p_3\simeq p_1^3\simeq 1.8\times 10^{-6}$.
This suggests that all sources that have three and more photons associated with them are individually detected above 100~GeV at $\ge 4\sigma$ level. The probability to find a photon within a $0.4^\circ$ circle (about the size of the 95\% containment radius of the PSF) is $\tilde p_1=0.195$. The probability to find 2 photons within the 68\% containment radius and an additional photon within the 95\% radius is $\simeq 3\times 10^{-5}$, also below the $4\sigma$ level. In a similar way, the probability to fine one photon within the 68\% containment circle and four additional photons within the 95\% containment circle is $\simeq 2\times 10^{-5}$, below the $4\sigma$ probability level.  We use these selection criteria to compile the list of sources detected in the VHE band, presented in Tables \ref{tab:known_sources} and \ref{tab:new_sources}. Table \ref{tab:known_sources} lists the sources that have already been reported in the VHE band based on observations with IACTs (sources present in the TeVCat compilation), while Table \ref{tab:new_sources} lists the sources that have not been previously reported as VHE emitters.  In total 53 out of 85 sources Table \ref{tab:new_sources} are detected by Fermi LAT with significance 5 $\sigma$ or higher.

Detection of 2 photons in the 68\% circle without additional photons within the 95\% circle corresponds to the chance coincidence probability $1.5\times 10^{-4}$ which is better than $3\sigma$ detection threshold. The probability to find 1 photons within 68\% radius and additional 2 photons within 95\% radius is $5\times 10^{-4}$, also below the $3\sigma$ level chance coincidence probability. Tables \ref{tab:known_supp} and \ref{tab:new_supp} list sources selected using these less strict criteria, with $>3\sigma$ detection significance, previously detected by IACTs and new detections. There are 5071 source in the 4FGL catalog in the sky region of interest and we expect 3 sources out of 100 listed in Tables \ref{tab:known_supp} and \ref{tab:new_supp} to be "false positives".  One source out of 71 with  2 photons within 68\% containment circle is a "false positive" and 2 sources out of 21 with  1 photon in the 68\% circle and additional 2 photons within the 95\% circle by chance are "false positives" in Tables \ref{tab:known_supp} and \ref{tab:new_supp}.  

Tables \ref{tab:known_sources}-\ref{tab:new_supp} list source fluxes in the VHE band. To derive the flux estimates, we use Fermi Science Tools (https://fermi.gsfc.nasa.gov/ssc/data/analysis/software/). We calculate LAT exposure in the source direction, using {\it gtselect-gtmktime-gtltcube-gtexpmap} tool chain as described in the Analysis Threads provided by the Fermi Science Support Center (https://fermi.gsfc.nasa.gov/ssc). We also present spectra of selected sources. The spectra are extracted using the aperture photometry method by counting events within 68\% ($C_{68}$) and 95\% ($C_{95}$) containment circles in each energy bin and estimating the source counts $S=(R_{95}^2C_{68}-R_{68}^2C_{95})/(0.68R_{95}^2-0.95R_{68}^2)$ where $R_{68}$ and $R_{95}$ are the 68\% and 95\% containment radii of the PSF in the energy range below 50~GeV and $S=C_{95}/0.95$ at higher energies where the background level is much lower than one count per square degree. We have verified that such approach provides flux measurements compatible with those reported in 4FGL.

\section{Properties of the VHE AGN population}

\subsection{Catalog content and completeness}

Fermi/LAT telescope continuously performs all sky survey, so that the list of 275 sources presented in  Tables
\ref{tab:known_sources}-\ref{tab:new_supp} provides a sky survey not biased by selection effects. The mean Fermi/LAT exposure over the sky is $AT=6\times 10^{11}$~cm$^2$s. On average, a source with flux $F=10^{-12}$~erg/(cm$^2$s) at $E=100$~GeV is expected to be found with $N_{ph}=F\cdot AT/E\simeq 3.75$ photons within Fermi/LAT exposure. This photon statistics is comparable to that of sources found in Tables \ref{tab:known_sources}-\ref{tab:new_supp}. Nevertheless, due to statistical fluctuations, it can happen that none of the photons associated to the source is found within the 68\% containment radius of the PSF or there is one such photon and the overall signal statistics is two photons. In this case the source would not be listed in the tables. If  $N_{ph}=3.75$, this happens in 20\% of cases (we perform small Monte-Carlo simulaitons to do the estimate).  We estimate our VHE AGN catalog  becomes 90\% complete for sources providing $N_{ph}=4.8$ photons on average. This corresponds to the flux $F=1.3\times 10^{-12}$~erg/cm$^2$s. 

\begin{figure}
    \includegraphics[width=\columnwidth]{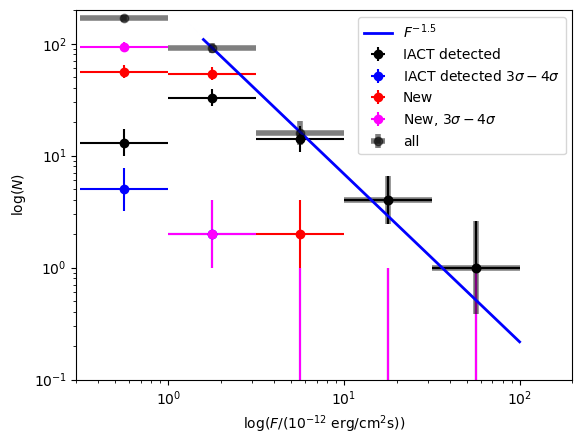}
    \caption{Number of sources as a function of flux. Black and blue data points are sources previously detected with IACTs. Black shows sources detected with $>4\sigma$ significance by Fermi/LAT in the VHE band, blue is for the sources with $3\sigma-4\sigma$ detection significance. Red and magenta points are new detections, with red color showing $>4\sigma$ sources and magenta the $3\sigma-4\sigma$ sources. Blue line shows the $F^{-3/2}$ scaling. }
    \label{fig:logN_logS}
\end{figure}

Fig. \ref{fig:logN_logS} shows the source counts $N$ as a function of their flux $F$. One can see that the $\log(N)-\log(F)$ distribution is consistent with the $F^{-3/2}$ scaling expected for the complete source sample. Fig. \ref{fig:logN_logS}  also shows contributions of sources from different tables into the overall source count statistics. Sources previously detected by IACTs have on average higher flux. New high significance detections (red data points are complementing the previous detections to yield the expected $F^{-3/2}$ scaling. Finally, lower significance sources, represented by magenta and blue data points, have lower flux and the deviation from the $F^{-3/2}$ indicates that the catalog is not complete at the flux level below $10^{-12}$~erg/cm$^2$s. 

\begin{figure}
    \includegraphics[width=\columnwidth]{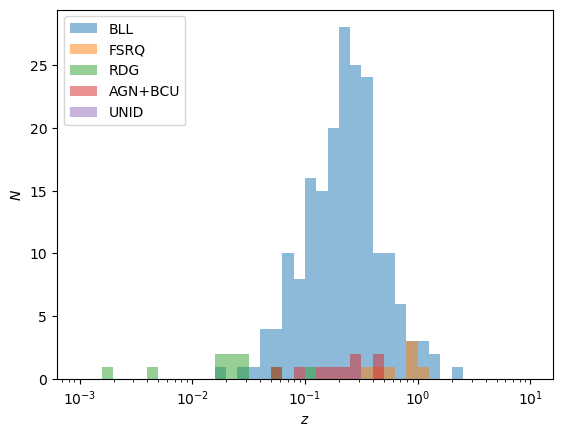}
    \caption{Number of sources as a function of redshift. }
    \label{fig:redshifts}
\end{figure}

Fig. \ref{fig:redshifts} shows source distribution as a function of redshift extracted from SIMBAD astronomical data service (https://simbad.u-strasbg.fr/). This figure also provides information on statistics of source detections by source type. We use source classification from the Fermi catalog \citep{Ballet:2023qzs}. Most of the sources (233) are BL Lacs (BLL), which is the dominant VHE AGN type at the survey sensitivity limit. The second-largest sample is that of Radio Galaxies (RDG), which consists of 10 sources. Seven sources are Flat Spectrum Radio Quasars (FSRQ). One source (M82) is a Starburst Galaxy (SBG).  20 AGN in the sample are not classified (their class is "AGN" in the Fermi catalog), and/or considered as "blazar candidates" ("BCU" in the Fermi catalog). Some of the sources listed as "BCU" in the 4FGL catalog have been found to belong to the BL Lac or FSRQ classes \citep{2022MNRAS.516.5702O} and we include this information in our source list. 

The catalog presented in Tables \ref{tab:known_sources}-\ref{tab:new_supp} does not include sources that are explicitly not AGN. Several such sources are present in the part of the sky considered in  our analysis, such as sources in the Large Magellanic Cloud and PSR J1023+0038. Nevertheless, there are four sources that do not have identified counterparts in Fermi catalog. These sources may or may not be AGN and we include these sources in the list for completeness. 

\begin{figure}
    \includegraphics[width=\columnwidth]{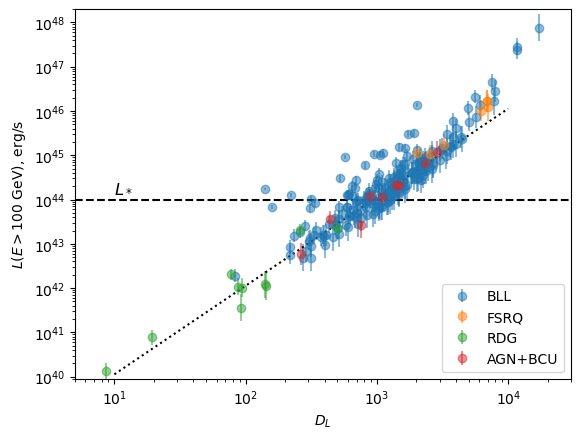}
    \caption{Intrinsic source luminosity as a function of distance. Dotted line shows the luminosity corresponding to the flux $F=10^{-12}$~erg/cm$^2$s. }
    \label{fig:distance_luminosity}
\end{figure}

\subsection{VHE AGN luminosity functions}
Fig. \ref{fig:distance_luminosity} shows the estimates of luminosities of sources as a function of distance, calculated using Planck 2018 cosmological parameters \citep{2020A&A...641A...6P}. Radio galaxies are found at smaller distances because they are intrinsically less bright than BL Lacs and FSRQs, with luminosities  in the $10^{40}-10^{44}$~erg/s range.  The bulk of BL Lacs is found at Gpc scale distances, with luminosities in excess of $10^{42}$~erg/s. Remarkably, the luminosities of the furthest away BL Lacs sources reach above $10^{47}$~erg/s, comparable or higher than the luminosity of the FSRQs detected in the VHE band. The luminosity shown in Fig. \ref{fig:distance_luminosity} is the intrinsic source luminosity corrected for the effect of attenuation of the \gr\ flux due to absorption in interactions with the EBL photons. We have used the model of \cite{2021MNRAS.507.5144S} to perform this correction.  From Fig. \ref{fig:redshifts} and \ref{fig:distance_luminosity} one can see that a number of "unclassified" AGN, "blazar candidates" and "unidentified" 4FGL sources have redshift estimates and luminosities in the range typical for BL Lacs and it is thus probable that those sources are from BL Lac class or, otherwise, they can belong to the population of radio galaxies.  Dotted line in Fig.  \ref{fig:distance_luminosity} shows the luminosity level corresponding to the limiting flux of our survey, $\simeq 10^{-12}$~erg/cm$^2$s. One can see that large number of sources is found along this line, which is in agreement with expectation from the $\log(N)-\log(F)$ distribution of Fig. \ref{fig:logN_logS}. 

\begin{figure}
    \includegraphics[width=\columnwidth]{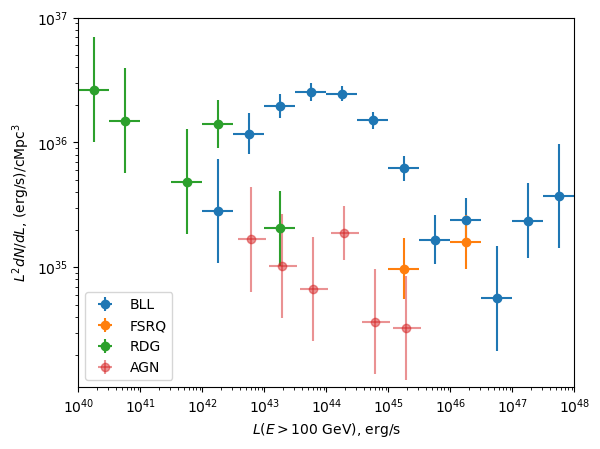}
    \caption{Luminosity functions of VHE AGN sub-classes. }
    \label{fig:luminosity_function}
\end{figure}

Fig. \ref{fig:luminosity_function} shows the luminosity functions of BL Lacs, FSRQs and radio galaxies detected in the VHE range (source density per comoving Megaparsec, cMpc, as a function of luminosity). The statistics of detections of radio galaxies and FSRQs is barely sufficient for extraction of the luminosity function. To the contrary, the luminosity function of BL Lacs is well measured and it can be modeled with a broken powerlaw function, with a change in the slope of the powerlaw $dN/dL\propto L^{-\kappa}$ from  $\kappa<2$ to $\kappa>2$ at the luminosity $L_*\sim 10^{44}$~erg/s, as it is clear from \ref{fig:luminosity_function}.  The overall power injection from the BL Lac population ($\propto\int LdN/dL dL\propto L^{2-\kappa}$) is dominated by sources with luminosities of the order of $L_*$. The deficit of source count statistics at $L<L_*$ is not due to the sensitivity limit of the survey, as it is clear from Fig. \ref{fig:distance_luminosity}: sources with luminosity order-of-magnitude lower than $L_*$ would be visible at the distances of closest VHE BL Lacs ($\sim 100$~Mpc) if they would be present in amounts larger of equal to those of sources with $L\sim L_*$ in this distance range. Remarkably, Fig. \ref{fig:luminosity_function} shows that the total VHE \gr\ power of the radio galaxy population may be comparable to that of BL Lacs, although low statistics of detections of radio galaxies prevents characterization of the radio galaxy luminosity function in the VHE band.    The total power of the VHE emitting BL Lacs, $W=(1.41\pm 0.11)\times 10^{37}$~erg/s/Mpc$^3$, is given by the integral of the luminosity weighted luminosity function. The energy density of \gr\ background generated by the BL Lac population over a cosmological time scale $T\sim 10^{10}$~yr can be estimated as $\rho=WT\simeq 10^{-7}$~eV/cm$^3$, which corresponds to the flux $F=\rho c/(4\pi)\simeq 2.3\times 10^{-4}$~MeV/cm$^2$s, at the level of the Extragalactic Gamma-Ray Backgorund (EGRB) measured by Fermi/LAT at 100~GeV energy \citep{2015ApJ...799...86A}.

\subsection{Blazar sequence and extreme sources}

\begin{figure}
    \includegraphics[width=\columnwidth]{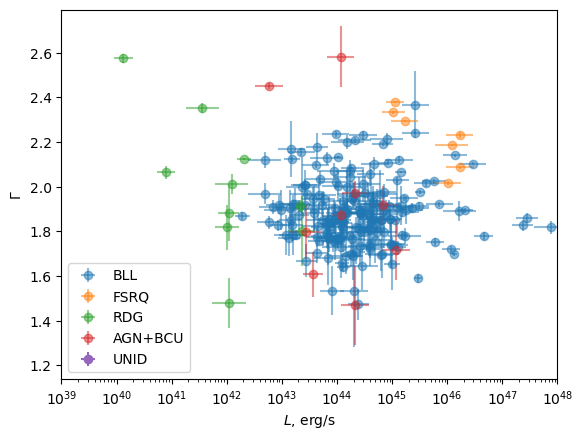}
    \caption{Spectral slope vs. VHE band luminosity for different source types.}
    \label{fig:HR}
\end{figure}

Most of the sources in the VHE AGN sample are blazars, that are known to have their spectral characteristics varying as a function of luminosity in the VHE band. This phenomenon is known as "blazar sequence": higher luminosity sources tend to have softer spectra \citep{1998MNRAS.299..433F,2022Galax..10...35P}. Fig. \ref{fig:HR} shows the slope of the spectrum as a function of source luminosity. The spectral slopes are from the Fermi/LAT catalog \citep{Ballet:2023qzs}, while the luminosity is estimated based on source redshifts and corrected for the effect of absorption of \gr s in interactions with EBL. In accordance with the expectations of the blazar sequence, the FSRQ source sample that has higher average luminosity compared to the BL Lac sample is also characterized by softer spectral slopes but within the BL Lac source sample sources with luminosity comparable or larger than those of the FSRQ can have harder spectra, so that there is no clear "blazar sequence" evident in the VHE band. It is possible that there exists a population of FSRQs with soft spectra and high luminosities, but these  FSRQs are not included in the present catalog because their VHE flux is below the flux limit of the catalog. 

Most of the sources from the BL Lac source sample have the spectral slopes within $\pm 0.1$ from the average slope  $\Gamma_*\simeq 1.8$. To the contrary, there is no well-defined slope of the radio galaxy source sample. Most of the unclassified AGN, blazar candidates and unidentified sources have the spectral properties consistent with those of the BL Lac sample: $L\sim L_*, \Gamma\sim \Gamma_*$, which suggests that they may be representatives of the BL Lac source class. 

The spectral energy distribution of VHE AGN typically has two components with the \gr\ component generated via inverse Compton scattering by energetic electrons and positrons, and a lower energy component produced via synchrotron emission by the same electrons. The spectral properties of the two components are correlated and it has been conjectured that the strongest VHE emitters are the "extreme" sources that have the peak of the synchrotron emission component in the X-ray band. A catalog of High Synchrotron Peak (HSP) sources has been compiled by  \citet{2019A&A...632A..77C}. We include information from this catalog in the last column of Tables \ref{tab:known_sources}-\ref{tab:new_supp} by tagging the extreme HSP sources with the "E" label. There are 22 such sources among previously reported VHE emitters (Tables \ref{tab:known_sources}, \ref{tab:known_supp}) and 41 among the new detections (Tables \ref{tab:new_sources},\ref{tab:new_supp}). 

\begin{figure}
    \includegraphics[width=\columnwidth]{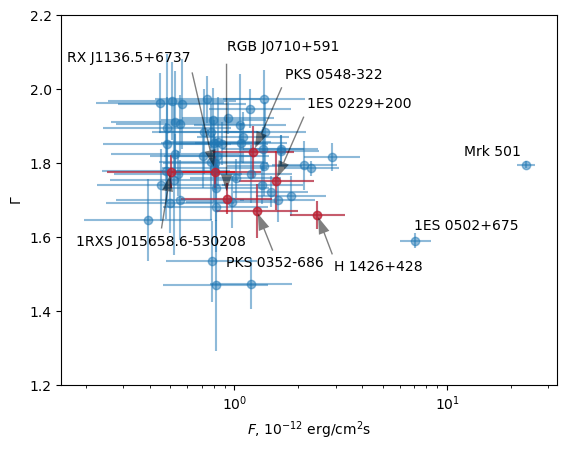}
    \caption{Spectral characteristics for the subset of "extreme" sources. Blue data points show sources with synchrotron spectrum peak at frequencies $\nu_s>10^{17}$~Hz. Red data points show sources with $\nu_s>10^{18}$~Hz.}
    \label{fig:extreme}
\end{figure}

\begin{figure*}
     \includegraphics[width=0.85\columnwidth]{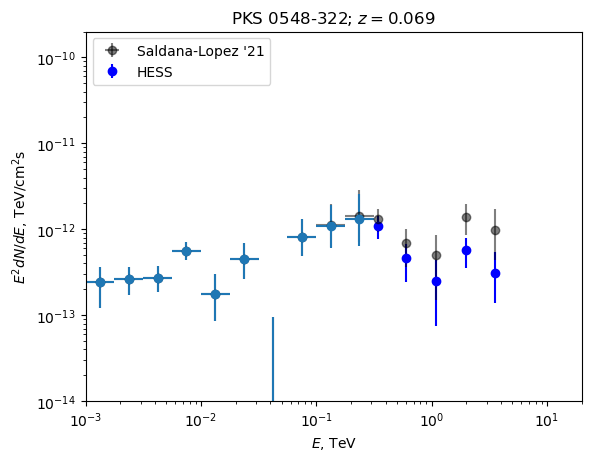}
    \includegraphics[width=0.85\columnwidth]{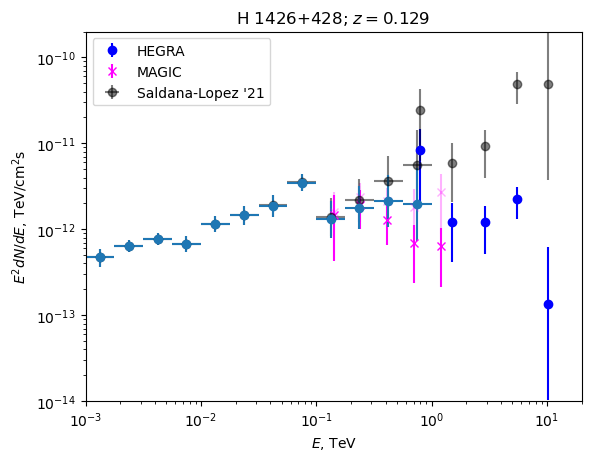}\\
    \includegraphics[width=0.85\columnwidth]{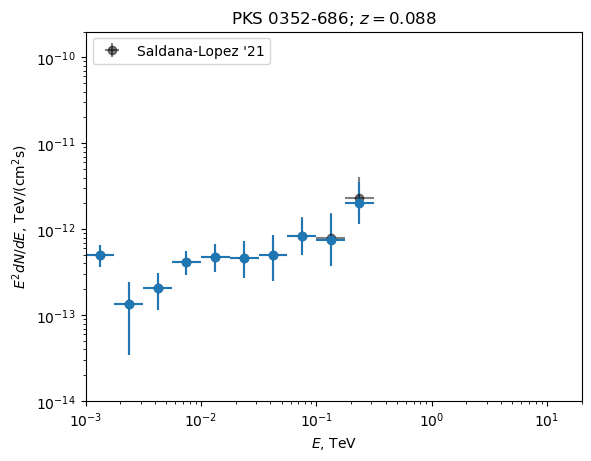}
    \includegraphics[width=0.85\columnwidth]{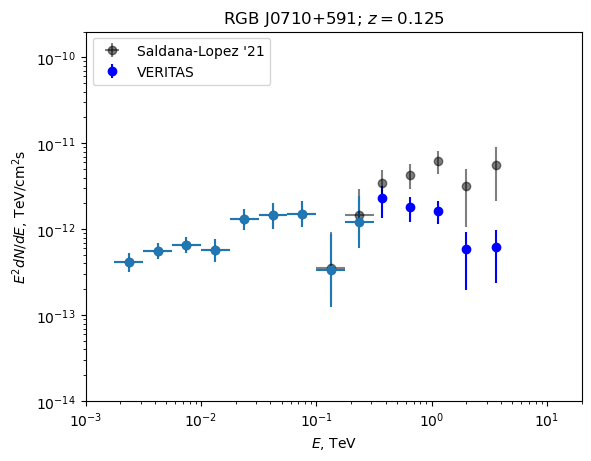}\\
    \includegraphics[width=0.85\columnwidth]{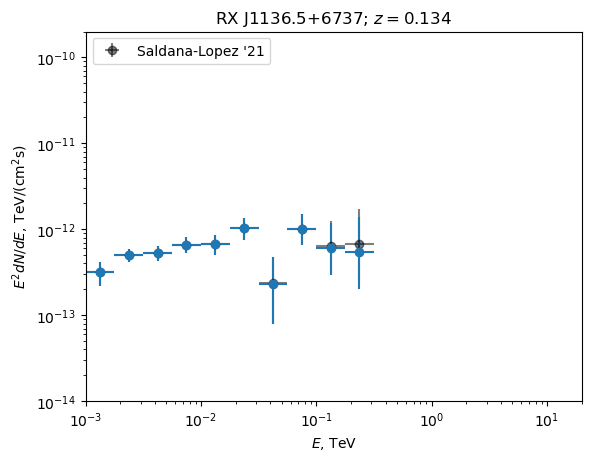}
     \includegraphics[width=0.85\columnwidth]{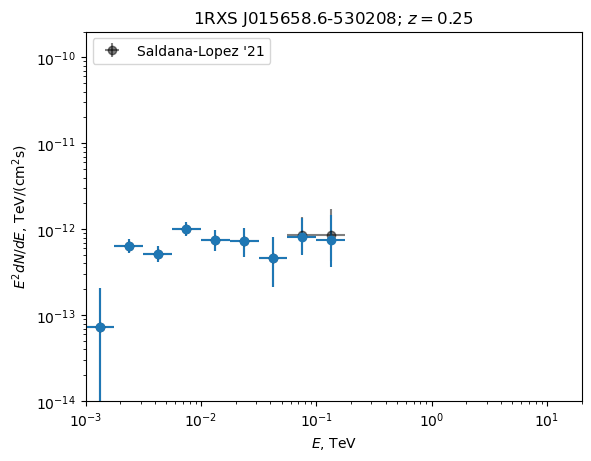}\\
    \includegraphics[width=0.85\columnwidth]{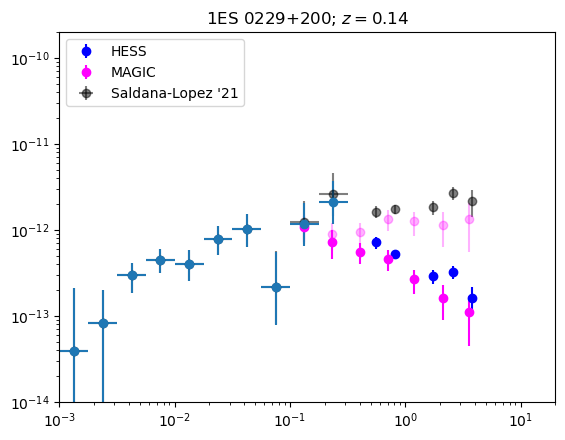}
\includegraphics[width=0.85\columnwidth]{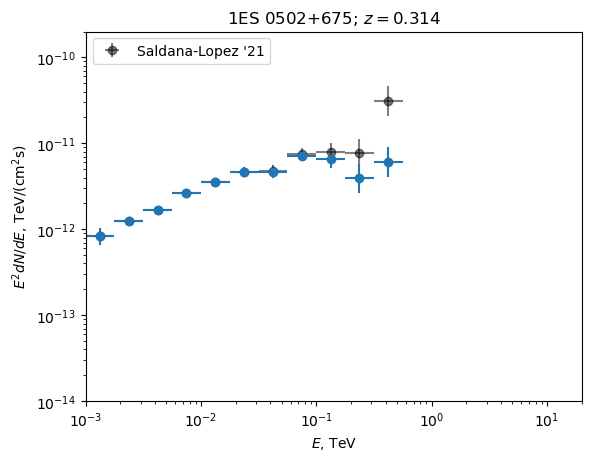}\\
   \caption{Spectra of the most extreme blazars (synchrotron spectrum cut-off in the $\nu_s>10^{18}$~Hz frequency range). The last panel shows the spectrum of 1ES 0502+675 that does not belong to the most extreme source sample, but is the second brightest source in the extreme VHE blazar sample.  }
    \label{fig:extreme_spectra}
\end{figure*}

The extreme blazars are particularly interesting in the context of the measurements of Intergalactic Magnetic Fields (IGMF) \citep{2010Sci...328...73N,2009PhRvD..80l3012N} via detection of secondary emission from electron-positron pairs deposited by the process of absorption of VHE \gr s on EBL in the intergalactic medium. The overall power of the secondary emission is determined by the source power at the highest energies, which is maximized in the extreme sources. Detection of the secondary signal in the form of extended emission superimposed on the primary point source signal is favored if the primary source spectrum is hard. Thus, the extreme VHE AGN with hard point source spectra provide the primary interest targets for the IGMF study. \cite{2010Sci...328...73N} have selected three extreme blazars that appeared the most promising candidates for the IGMF search based on the information on the VHE AGN sample available at the start of operations of Fermi/LAT. \citet{2023ApJ...950L..16A} have made a selection of sources based on the availability of data of HESS telescope. Both approaches are ad-hoc, rather than result of systematic analysis aimed at optimization of the secondary signal search. Better knowledge of the AGN VHE population achieved with the long LAT exposure enables a more systematic approach for selection of sources for the IGMF study. Fig. \ref{fig:extreme} shows the distribution of VHE fluxes and slopes of the \gr\ spectra of the extreme sources. The best candidates for the search of the secondary \gr\ signal are the sources with the highest VHE flux and hardest spectra extending to the highest energies.   The  highest flux sources apparent in Fig. \ref{fig:extreme} are Mrk 501 and 1ES 0502+675. 1ES 0502+675 also has hard spectrum (see Fig. \ref{fig:extreme_spectra}). Detection of 1ES 0502+675 in the VHE band has been reported by VERITAS, \citep{2011ICRC....8...51B}, but no publication with the details of spectral characteristics of the source is available so far.

Fermi/LAT cannot measure the extent of the spectrum in the multi-TeV range, but the energy of the synchrotron cutoff can serve as an indication of the TeV spectral cut-off. Red data points in Fig. \ref{fig:extreme} show sources that have the synchrotron cut-off frequency in excess of $10^{18}$~Hz (photon energy higher than $4$~keV). The source 1ES 0229+200 that provides tightest constraints on the IGMF \citep{2023A&A...670A.145A,2023ApJ...950L..16A} is one of such sources, but there are six more "the most extreme" blazars like 1ES 0229+200. H 1426+428 has been detected for the first time in the TeV band by HEGRA telescope \citep{2002A&A...384L..23A} and has been recently observed by MAGIC \citep{2020ApJS..247...16A}. RGB J0710+591 was observed by VERITAS \citep{2010ApJ...715L..49A}. Detection of RX J1136.5+6737 has been announced by MAGIC telescope \citep{2014ATel.6062....1M} back in 2014, but no publication with spectral characteristics of the source is available.  PKS 0548-322 has been detected by HESS \citep{2010A&A...521A..69A}. 1RXS J015658.6-530208 and PKS 0352-686 have not been observed by IACTs. Fig. \ref{fig:extreme_spectra} shows the spectra of the most extreme blazars detected in the VHE band. One can see that the spectra of the IACT detected sources are continuing toward the highest energies in the 10~TeV range as powerlaws without high-energy cut-offs, once corrected for the effect of absorption on EBL, as expected from the extremely high energy of the synchrotron spectrum cut-off. 

\subsection{Neutrino source candidates}

VHE emitting blazars, being the dominant source of the extragalactic \gr\ signal  \citep{Neronov:2011kg,2015ApJ...800L..27A},  are often considered as potential sources of VHE neutrinos. They are possibly a sub-dominant neutrino source class. Blazars detected in the GeV energy range can only provide a sub-dominant contribution to the astrophysical neutrino flux because the arrival directions of astrophysical neutrinos are not correlated with Fermi detected blazars \citep{Neronov:2016ksj,IceCube:2016qvd}. The VHE AGN catalog provides a different source selection compared to the blazar catalog derived from 4FGL and it is interesting to explore if the neutrino arrival directions correlate with the positions of the VHE detected AGN, or with a specific sub-class of the VHE detected AGN (FSRQ sources with high column density target material may be stronger neutrino sources compared to BL Lacs \citep{Neronov:2002xv}). 

The VHE AGN catalog includes sources that show evidence for neutrino emission, such as TXS 0506+056 \citep{2018Sci...361.1378I,IceCube:2018cha}. \cite{Padovani:2019xcv} have shown that TXS~0506+056 is special type of object, "masquerading BL Lac", namely an FSRQ with the emission lines heavily diluted by a strong Doppler-boosted jet. Another VHE blazar considered as neutrino source candidate is PKS 1424+240 \cite{Padovani:2022wjk}. Still one more source from the VHE AGN catalog, PKS 0735+178, number 37 in Table 2 (sources not yet observed with IACTs), was conjectured to be producing neutrinos emission correlated with \gr\ flaring activity  \citep{2023ApJ...954...70A,Sahakyan:2022nbz}. Other blazars from Table \ref{tab:known_sources}: 3C 66A, Mrk 421, PG 1553+113, 1ES 1959+650, Mrk 501, S5 0716+714, BL Lac, S4 0954+65, and Table \ref{tab:known_supp}: OJ 287, were considered as possible flaring neutrino sources by \citet{Oikonomou:2019djc}.

\subsection{High redshift sources}

Most remarkably, the VHE AGN catalog includes sources at redshifts $z\sim 1$ in spite of the fact that the VHE \gr\ flux from distant sources is strongly attenuated by the interactions with EBL photons. The optical depth with respect to the absorption on EBL reaches $\tau\simeq 1$ for the 100~GeV photons coming from redshift $z\simeq 1$ \citep{2017A&A...603A..34F}. This decreases statistics of the VHE signal and complicates detection of sources. The most distant source that has been detected in the VHE band up to now by IACTs in the VHE band is the blazar OP 313 at the redshift just below 1 (source number 39 in Table \ref{tab:known_sources}). Tables \ref{tab:known_sources}-\ref{tab:new_supp} contain seven sources with redshifts $z>1$, inferred from SIMBAD database (https://simbad.u-strasbg.fr/). Below we summarize the properties of these high-redshift VHE sources.

\begin{figure*}
     \includegraphics[width=0.85\columnwidth]{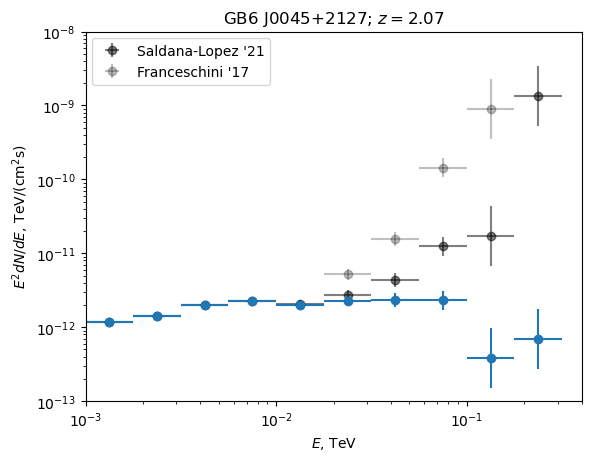}
    \includegraphics[width=0.85\columnwidth]{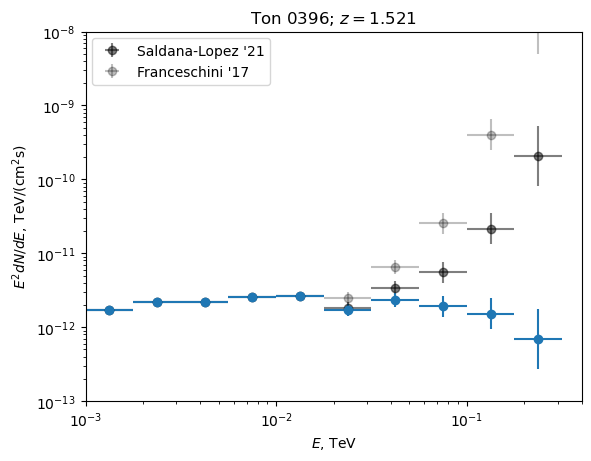}\\
    \includegraphics[width=0.85\columnwidth]{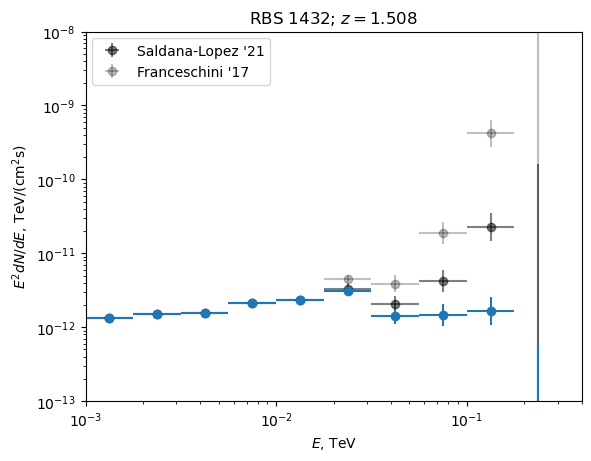}
    \includegraphics[width=0.85\columnwidth]{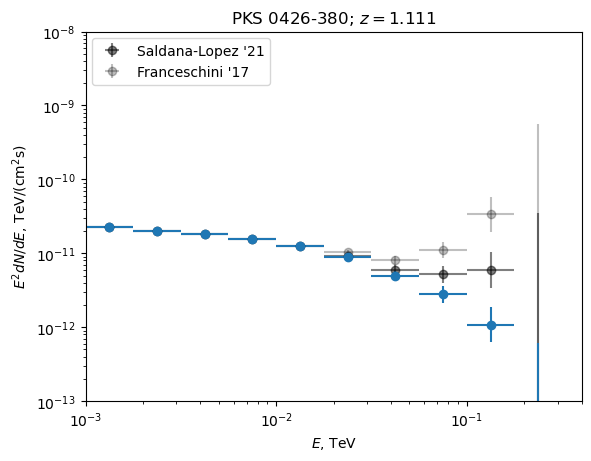}\\
    \includegraphics[width=0.85\columnwidth]{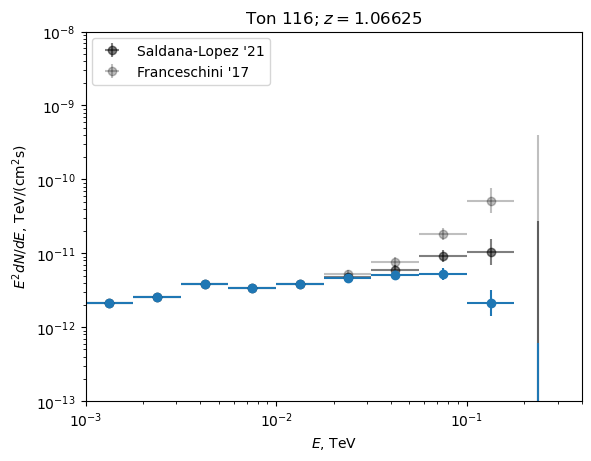}
     \includegraphics[width=0.85\columnwidth]{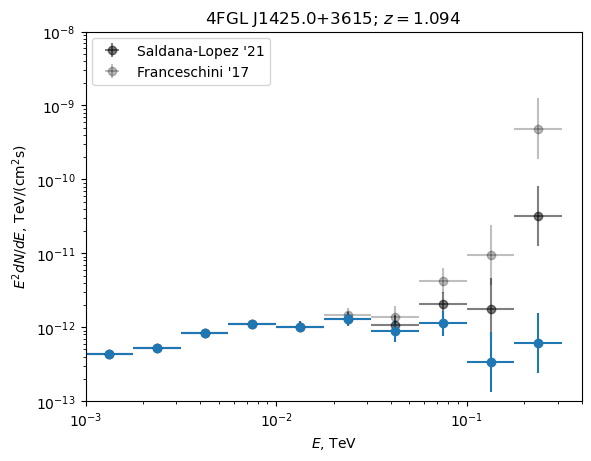}\\
    \includegraphics[width=0.85\columnwidth]{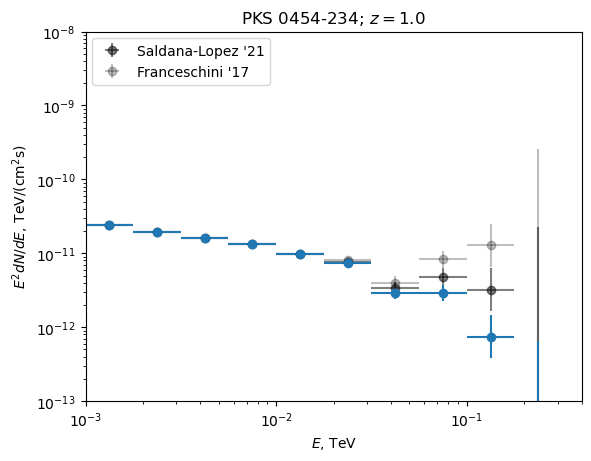}
   \caption{Spectra of high redshift sources. Black and gray data points show spectra corrected for EBL absorption based on two different models of EBL, specified in the figure legends. }
    \label{fig:high_redshift}
\end{figure*}

\textit{GB6 J0045+2127.}
This source (number 4 in Table \ref{tab:new_supp}) is formally the highest redshift source in the sample. SIMBAD provides the source redshift $z=2.07$ based on the  Sloan Digital Sky Survey  Data Release 16 spectroscopic data \citep{2020ApJS..250....8L}. The SDSS visual inspection redshift is judged to be good quality determination (2 out of 3 confidence score).  redshift has been derived based on a visible inspection of the source spectrum, rather than based on an automatic pipeline. The SDSS redshift determination by visual inspection has good quality. \cite{2020MNRAS.497...94P} have previously identified a faint Ca II absorption line doublet identified at redshift $z=0.4253$ in the source spectrum, which provides a lower bound on the source redshift, consistent with the SDSS measurement. To the contrary, the analysis of \cite{Shaw:2013pp} estimated  $z<1.78$ based on the absence of Lyman-alpha absorption features in the source spectrum. This estimate is in tension with the SDSS measurement. Finally, \cite{2024A&A...691A.154N} used a method of masking the blazar signal to detect the host galaxy of the source and estimate the redshift $z\simeq 0.34$ based on the imaging properties of the host galaxy.

The first panel of the first row of Fig. \ref{fig:high_redshift} shows the source spectrum corrected for the EBL absorption using two alternative EBL models: of \cite{2017A&A...603A..34F} and \cite{2021MNRAS.507.5144S}. In both cases, the absorption corrected source spectrum shows a peculiar hardening above 100~GeV, which is not found in low-redshift sources and is likely the artifact of wrong correction for the EBL absorption. This can be because the redshift measurement from SDSS DR16 \citep{2020ApJS..250....8L} is wrong, or because the EBL "nominal" models are over-estimating the density of the EBL at redshifts $z\gtrsim 1$. In the particular case of the model of \citet{2021MNRAS.507.5144S}, the hardening of the spectrum can be avoided if the optical depth with respect to absorption on the EBL is reduced by a factor of $\simeq 2$. The factor-of-two uncertainty  of the EBL spectrum normalization at $z\sim 2$ is well within the error budget of the models of \cite{2021MNRAS.507.5144S} as can be seen from Fig. 12 of that paper. A stronger suppression of the optical depth (reduction of normalization of the EBL spectrum) is needed in the case of \citet{2017A&A...603A..34F} model. 

\textit{Ton 0396}
 (source number 45 in Table \ref{tab:new_sources})  has redshift $z=1.52117$ in SIMBAD, based on SDSS Data Release 7 \cite{SDSS:2008tqn}. However, a study of \citet{2017ApJ...837..144P} finds a featureless spectrum of the source and describes the source as being with unknown redshift. An upper limit on the source redshift from spectroscopic study of ref.  \cite{Shaw:2013pp} $z<1.69$ is consistent with the SDSS value. The source has a relatively high VHE photon statistics with 5 photons in 95\%  containment radius. Similar to GB6 J0045+2127, its spectrum, shown in Fig. \ref{fig:high_redshift} shows a pronounced hardening in the VHE range, either because the redshift of the source from the SDSS is not correct or because of the over-estimate of the optical depth with respect to the absorption on EBL. Similar to the case of GB6 J0045+2127, a factor-of-two decrease of the optical depth by a factor of 2 in \cite{2021MNRAS.507.5144S} model removes the hardening. The factor of two is marginally consistent the model of \cite{2021MNRAS.507.5144S} given the systematic uncertainty of the EBL normalization at redshift 1.5.

\textit{RBS 1432.}
SIMBAD ascribes the spectroscopic redshift $z=1.508$ to this source based on the DR7 of SDSS \cite{SDSS:2008tqn}. At the same time, the source is listed in the SDSS DR16Q \citep{2020ApJS..250....8L} where it has  a lower redshift, $z=1.3801$, quoted together with a set of warnings about from the spectral analysis pipeline, concerning bad astrometry and negative line normalization fit. Spectroscopic study of  ref. \cite{Shaw:2013pp} show minimal and maximal limits on redshifts $z>0.738$   $z<1.76$ which are consistent with this value. The source is detected by Fermi LAT above 100 GeV with 5 photons, 4 of which are withing 68\% from source position. Total significance of detection is above 5 $\sigma$ at 100 GeV. The spectrum of the source shown in Fig. \ref{fig:high_redshift} shows a hardening in the VHE band, which is especially pronounced if the correction for the EBL absorption is done using \cite{2017A&A...603A..34F} model. Revising the source redshift toward the lower value suggested by the SDSS DR16Q analysis reduces the hardening. 

\textit{PKS 0426-380}
has redshift $z=1.110$ based on 2MASS survey \citep{2011NewA...16..503M}.  It is classified  as BL Lac in 4FGL catalog \citep{Ballet:2023qzs}, but in recent study of Blazar Broad Emission Lines of \cite{2022ApJ...936..146X} it was reclassified as an FSRQ. Contrary to higher redshift sources, the spectrum of PKS 0426-380 corrected for absorption on EBL using \cite{2021MNRAS.507.5144S} model does not show hardening in the VHE range which points to improvement of the quality of the EBL modeling toward redshifts $z\sim 1$.

\textit{Ton 116}
is a BL Lac object with redshift  $z=1.06625$ in SIMBAD database, based on the SDSS DR7 measurement \cite{SDSS:2008tqn} but it was reconsidered to $z=1.182073$ SDSS Data Release 13 \citep{2017ApJS..233...25A}.  \cite{Shaw:2013pp} derive lower and upper limits on redshift $z>0.485$ and $z<1.77$.
In dedicated study of 10 m Gran Telescopio Canarias \cite{Paiano:2017pol} only a low limit on redshift was established $z>0.483$.  This source already had 100 GeV photon in first observations by Fermi LAT and  was discussed in \cite{2015A&A...575A..21N}. Today it detected by Fermi with significance well above 5 $\sigma$ above 100 GeV with 4 photons in 68 \%  and two additional photons in 95\% containment circles of the PSF (see Table 2). Fig. \ref{fig:high_redshift} in 3rd row left shows the source spectrum which is consistently a simple hard powerlaw, once corrected for the EBL absorption using the model of  \cite{2021MNRAS.507.5144S}. The spectrum is still hardening above 100 ~GeV if the EBL model of  \cite{2017A&A...603A..34F} is used.

\textit{4FGL J1425.0+3615}
 is a BL Lac object with redshift $z=1.09415$ based on the SDSS DR7 measurement \citep{SDSS:2008tqn}.
This redshift is consistent with the upper limit $z<2.17$ from spectroscopic study of ref.\cite{Shaw:2013pp}.  Similar to Ton 116, the source spectrum, shown in the right column, 4th row of Fig. \ref{fig:high_redshift}, is a hard powerlaw, when corrected for the effect of the EBL absorption using \cite{2021MNRAS.507.5144S} model. Remarkably, the highest energy photon associated to the source has energy in excess of 200~GeV.

\textit{PKS 0454-234}
is an FSRQ object  at redshift $z=1.003$ as found by \cite{Paliya:2021tfi} based on detection of MgII emission line. .  The source spectrum shown in Fig. \ref{fig:high_redshift} is a soft powerlaw, when corrected for the EBL absorption with either \cite{2021MNRAS.507.5144S} or \cite{2017A&A...603A..34F} model.

\subsection{Unidentified sources}

Four sources in Table \ref{tab:new_supp} do not have identified counterparts, but they can still be distant AGN or they can even be "echos" of past AGN activity of some galaxies, produced by the secondary \gr\ emission from VHE \gr s propagating through the intergalactic medium and interacting with the EBL. Fig. \ref{fig:unids} shows the spectra of these sources. The source spectra are consistently hard, which is promising for their detection with IACTs, but the source fluxes in the VHE band are all close to $10^{-12}$~erg/cm$^2$s, which is the sensitivity limit of CTAO for a nominal 50~hr exposure. 

\textit{4FGL J0032.3-5539} only has several galaxies, not identified as AGN, within $0.1^\circ$ from the 4FGL source position. Otherwise, another 4FGL source, tentatively identified with a radio source SUMSS J003210-552228 is found within the 95\% containment circle of the PSF. The other source is brighter than 4FGL J0032.3-5539 in the GeV band, but the VHE emission is clearly associated to 4FGL J0032.3-5539. 

\begin{figure}
    \includegraphics[width=\columnwidth]{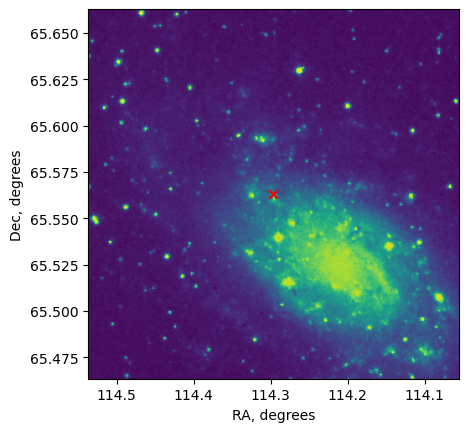}
    \caption{SDSS image of the field (downloaded from SkyView service (https://skyview.gsfc.nasa.gov/) around 4FGL J0737.4+6535 showing the VHE source position (red marker) within the nearby galaxy NGC 2403. }
    \label{fig:ngc_2403}
\end{figure}

\begin{figure*}
     \includegraphics[width=0.85\columnwidth]{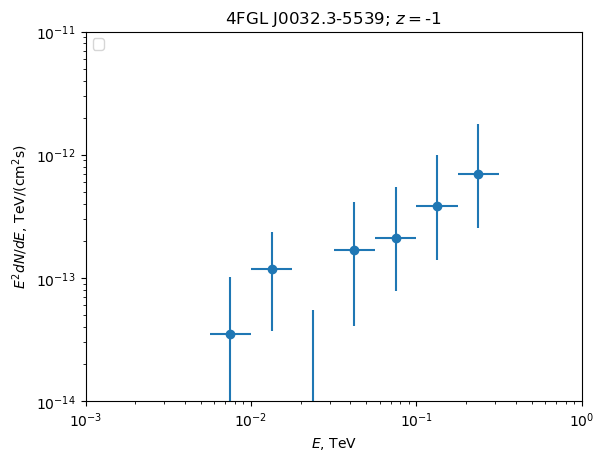}
    \includegraphics[width=0.85\columnwidth]{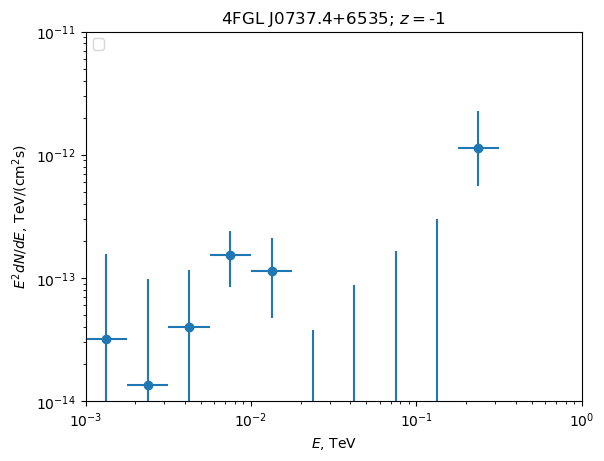}\\
     \includegraphics[width=0.85\columnwidth]{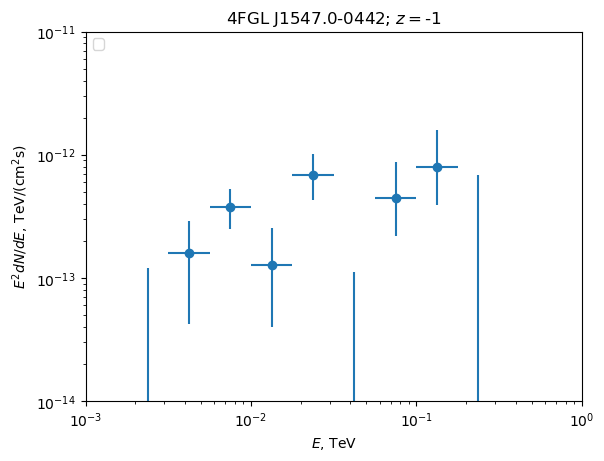}
    \includegraphics[width=0.85\columnwidth]{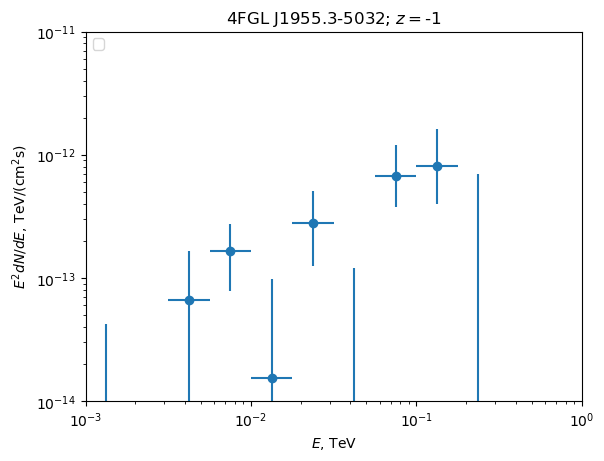}
   \caption{Spectra of unidentified 4FGL sources visible in the VHE band. }
    \label{fig:unids}
\end{figure*}

\textit{4FGL J0737+6535} is situated within the extent of the disk of NGC 2403 at 2.6 Mpc distance from the Milky Way. The source is displaced from the center of the galaxy, so that is is hardly an AGN. Several supernova remnants and X-ray binaries can be considered as counterparts of this source (Fig. \ref{fig:ngc_2403}). Similar to  4FGL J0032.3-5539, the source shows hard spectrum with two photons above 200~GeV associated to it (Fig. \ref{fig:unids}). Given the distance to the source, the luminosity of the object in NGC 2403 responsible for the VHE emission is exceptionally high, $\sim 10^{39}$~erg/s, exceeding the luminosity of the brightest sources in the Milky Way, like Crab Nebula.

\textit{4FGL J1547.0-0442} does not have obvious counterpart candidates, except for a radio source NVSS J154713-044415 3 arcmin away. The source spectrum (Fig. \ref{fig:unids}) is not as hard as that of 4FGL J0032.3-5539 and 4FGL J0737+6535, but still its slope is harder than $\Gamma=2$ which favors its detection at higher energies with IACTs. 

\textit{FGL J1955.3-5032} is listed as "unidentified" source in the 4FGL catalog, and there is only one galaxy, not classified as an AGN, within $0.1^\circ$ distance from the source. \citet{2021ApJ...923...75K} have put forward a hypothesis that the source is a blazar candidate, based on its spectral characteristics, using machine learning technique to find similarities of the source with known blazars. In such settings, a new source type, e.g. "blazar remnant" produced by secondary emission from \gr\ beam interactions in the intergalactic medium, would not be discovered.

\textit{4FGL J2317.7+2839} has several quasar counterpart candidates within $0.1^\circ$ from its catalog position, but no clear identification has been proposed so far. Similar to FGL J1955.3-5032, machine learning technique has been applied to this source, to search for similarities with the blazar characteristics  \citep{2021ApJ...923...75K}.

\section{Discussion and conclusions}

Our analysis of long exposure of Fermi/LAT toward extragalactic sky $|b|>10^\circ$ reveals a set of 175  AGN seen with significance above 4 $\sigma$ in VHE band (at energies $E>100$ GeV) and additionally 100 AGN seen with significance above 3$\sigma$ in VHE band. The VHE AGN catalog content is dominated by BL Lac type objects (233), but includes a limited number of radio galaxies (10), FSRQs (7) and one starburst galaxy. Also there are 20 unidentified type of blazars (BCU) and 4 unidentified objects. The catalog is complete at 90\% down to the flux limit of $1.3\times 10^{-12}$~erg/cm$^2$s. Only a minor part of the catalog (71 sources) lists previously reported to be VHE emitters, they are presented in the Table  \ref{tab:known_sources} for 4$\sigma$ sources and Table \ref{tab:known_supp} for 3$\sigma$ sources.  Majority of sources (204) are new detections listed in Table \ref{tab:new_sources}  for 4$\sigma$ sources and  Table \ref{tab:new_supp} for 3$\sigma$ sources.

The flux limit of the catalog is comparable to the sensitivity of the next-generation IACT facility, CTAO, that is in construction now. This means that the catalog presented in this paper can serve as a reference catalog for CTAO: most of the sources listed in the catalog are detectable with CTAO except for possibly sources that would be observable at large zenith angles from the Northern and Southern CTAO sites. The catalog  can be use to optimize planning of pointed observations with CTAO and with current generation IACTs, for the purpose of a VHE band extragalactic sky survey with relatively short exposures per source or for selection of best sources for deep exposures needed for the studies aimed at characterization of the EBL and IGMF. The VHE AGN catalog also includes blazars for which evidence for neutrino signal has been previously reported.

Specifically useful for the EBL and IGMF studies are "extreme" sources that are expected to have their \gr\ spectra peaking well in the multi-TeV range, because their synchrotron spectra are found to peak in the frequency range in excess of $10^{17}$Hz.  The VHE AGN catalog contain 49 such extreme sources. Their spectral index in Fermi LAT and flux in VHE band are presented in the  Fig. \ref{fig:extreme}. Out of those sources 7 have highest synchrotron peak    $\nu > 10^{18}$~Hz. Their spectral characteristics are presented in  Fig. \ref{fig:extreme_spectra}.  Only one of these sources, $1ES~0229+200$, has been a specific focus of deep exposures with IACTs so far. Other three sources, PKS~0548-322, H~1426+428, RGB~J0710+591, have been detected by IACTs, but not subject to detailed spectral characterization with deep exposures. Three sources, PKS~0352-686, RX~J1136.5+6737 and 1RXS~J015658.6-530208 have never been observed with IACTs. 

Our catalog includes  seven sources at redshift above $z=1$. Such sources have never been observed with IACTs, possibly because it is not expected that such sources are detectable. Fermi/LAT measurements of their flux in the energy range above 100~GeV assure that these sources are detectable with IACTs. Analysis of the Fermi/LAT spectra of these sources shows that the most-recent state-of-art EBL models by \citet{2021MNRAS.507.5144S} provide a reasonable description of the effect of attenuation of the \gr\ flux up to the redshift $z\sim 1$ (for sources PSK 0454-234, Ton 116, 4FGL J1425+3615, PKS 0426-380, see Fig. \ref{fig:high_redshift}). To the contrary, the spectra of sources well beyond $z=1$ show peculiar hardenings that indicate limitations of the EBL models at higher redshifts (RBS 1432, Ton 0396, GB6 J0045+2127). Such hardening can be avoided if the level of the EBL flux is reduced by a factor of two at high redshift, which is within the systematic uncertainty of the current state-of-art EBL models. The high redshift VHE AGN deserve special focus with deep IACT observations and improved optical spectroscopy to verify their redshift estimates.  Higher statistics measurements of the spectra of high redshift sources with IACTs can be used to improve understanding of cosmological evolution of the EBL and for the measurement of the Hubble constant \citep{Dominguez:2023rxa}, as well as for the measurements of  IGMF \citep{2010Sci...328...73N} and for the studies of constraints on Beyond-Standard-Model phenomena that may be revealed in the VHE \gr\ propagation over cosmological distances   \citep{2019ICRC...36..739M}.

The VHE AGN catalog presented here can serve as a reference catalog exploration of the high latitude  sky ($|b|>10^\circ$)   with existing and new generation and IACT systems: CTAO \citep{2019scta.book.....C} and Large Area Cherenkov Telescopes (LACT) \citep{2025ChPhC..49c5001Z}. Our catalog has advantage as compared to standard CTA approach of \cite{CTAConsortium:2013tmf} based on extrapolation of spectra of sources from $E<100$ GeV energies to the VHE band. It allows to avoid uncertainty of possible intrinsic cutoffs in  spectra of some sources. It can be an important tool for optimization of planning of IACT observations, reducing the loss of observation time on sources that do not have sufficiently high VHE band flux.  Given that detection of a source with the flux $\sim 10^{-12}$~erg/cm$^2$s at 100~GeV takes $\sim 50$~hr of CTAO observation time, the cumulative exposure of CTAO aimed at an AGN survey down to this flux limit would talk $275$ (sources in the VHE AGN catalog) $\times 50$~hr~$\sim 1.4\times 10^4$~hr, which is the observation time available with CTAO only on a decade time span, even if no other observation, apart from the extragalactic sky survey is performed. 

Apart from Fermi/LAT, air shower arrays like HAWC \citep{2021ApJ...907...67A} and LHAASO \citep{2024ApJS..271...25C} provide unbiased Northern sky surveys in the TeV energy range and in future the Southern Wide-field Gamma-ray Observatory (SWGO) \citep{2019BAAS...51g.109H} will complement the TeV sky survey with the Southern Hemisphere exposure. We expect that a combination of the VHE survey performed with Fermi/LAT at 100~GeV energy with the TeV sky surveys with LHAASO and SWGO has a potential to crucially improve the knowledge of the VHE emitting AGN population, complementing pointed observations with IACTs that provide "snapshot" source spectra, rather than unbiased measurements of time averaged source fluxes.

\section{Acknowledgments}
This work has been supported in part by the French National Research Agency (ANR) grant ANR-24-CE31-4686. 

\bibliographystyle{aa}
\bibliography{refs}

\begin{thebibliography}{61}
\expandafter\ifx\csname natexlab\endcsname\relax\def\natexlab#1{#1}\fi

\bibitem[{Aartsen {et~al.}(2017)}]{IceCube:2016qvd}
Aartsen, M.~G. {et~al.} 2017, Astrophys. J., 835, 45

\bibitem[{Aartsen {et~al.}(2018)}]{IceCube:2018cha}
Aartsen, M.~G. {et~al.} 2018, Science, 361, 147

\bibitem[{Abazajian {et~al.}(2009)}]{SDSS:2008tqn}
Abazajian, K.~N. {et~al.} 2009, Astrophys. J. Suppl., 182, 543

\bibitem[{{Acciari} {et~al.}(2023){Acciari}, {Agudo}, {Aniello}, {Ansoldi},
  {Antonelli}, {Arbet Engels}, {Artero}, {Asano}, {Baack}, {Babi{\'c}},
  {Baquero}, {Barres de Almeida}, {Barrio}, {Batkovi{\'c}}, {Becerra
  Gonz{\'a}lez}, {Bednarek}, {Bernardini}, {Bernardos}, {Berti}, {Besenrieder},
  {Bhattacharyya}, {Bigongiari}, {Biland}, {Blanch}, {B{\"o}kenkamp},
  {Bonnoli}, {Bo{\v{s}}njak}, {Burelli}, {Busetto}, {Carosi}, {Ceribella},
  {Cerruti}, {Chai}, {Chilingarian}, {Cikota}, {Colombo}, {Contreras},
  {Cortina}, {Covino}, {D'Amico}, {D'Elia}, {da Vela}, {Dazzi}, {de Angelis},
  {de Lotto}, {Del Popolo}, {Delfino}, {Delgado}, {Delgado Mendez}, {Depaoli},
  {di Pierro}, {di Venere}, {Do Souto Espi{\~n}eira}, {Dominis Prester},
  {Donini}, {Dorner}, {Doro}, {Elsaesser}, {Fallah Ramazani}, {Fari{\~n}a},
  {Fattorini}, {Font}, {Fruck}, {Fukami}, {Fukazawa}, {Garc{\'\i}a L{\'o}pez},
  {Garczarczyk}, {Gasparyan}, {Gaug}, {Giglietto}, {Giordano}, {Gliwny},
  {Godinovi{\'c}}, {Green}, {Green}, {Hadasch}, {Hahn}, {Hassan}, {Heckmann},
  {Herrera}, {Hrupec}, {H{\"u}tten}, {Inada}, {Iotov}, {Ishio}, {Iwamura},
  {Jim{\'e}nez Mart{\'\i}nez}, {Jormanainen}, {Jouvin}, {Kerszberg},
  {Kobayashi}, {Kubo}, {Kushida}, {Lamastra}, {Lelas}, {Leone}, {Lindfors},
  {Linhoff}, {Liodakis}, {Lombardi}, {Longo}, {L{\'o}pez-Coto},
  {L{\'o}pez-Moya}, {L{\'o}pez-Oramas}, {Loporchio}, {Lorini}, {Machado de
  Oliveira Fraga}, {Maggio}, {Majumdar}, {Makariev}, {Mallamaci}, {Maneva},
  {Manganaro}, {Mannheim}, {Mariotti}, {Mart{\'\i}nez}, {Mas Aguilar}, {Mazin},
  {Menchiari}, {Mender}, {Mi{\'c}anovi{\'c}}, {Miceli}, {Miener}, {Miranda},
  {Mirzoyan}, {Molina}, {Mondal}, {Moralejo}, {Morcuende}, {Moreno}, {Moretti},
  {Nakamori}, {Nanci}, {Nava}, {Neustroev}, {Nievas Rosillo}, {Nigro},
  {Nilsson}, {Nishijima}, {Noda}, {Nozaki}, {Ohtani}, {Oka}, {Otero-Santos},
  {Paiano}, {Palatiello}, {Paneque}, {Paoletti}, {Paredes}, {Pavleti{\'c}},
  {Pe{\~n}il}, {Persic}, {Pihet}, {Prada Moroni}, {Prandini}, {Priyadarshi},
  {Puljak}, {Rhode}, {Rib{\'o}}, {Rico}, {Righi}, {Rugliancich}, {Sahakyan},
  {Saito}, {Sakurai}, {Satalecka}, {Saturni}, {Schleicher}, {Schmidt},
  {Schmuckermaier}, {Schubert}, {Schweizer}, {Sitarek}, {{\v{S}}nidari{\'c}},
  {Sobczynska}, {Spolon}, {Stamerra}, {Stri{\v{s}}kovi{\'c}}, {Strom},
  {Strzys}, {Suda}, {Suri{\'c}}, {Takahashi}, {Takeishi}, {Tavecchio},
  {Temnikov}, {Terzi{\'c}}, {Teshima}, {Tosti}, {Truzzi}, {Tutone}, {Ubach},
  {van Scherpenberg}, {Vanzo}, {Vazquez Acosta}, {Ventura}, {Verguilov},
  {Viale}, {Vigorito}, \& {Vitale}}]{2023A&A...670A.145A}
{Acciari}, V.~A., {Agudo}, I., {Aniello}, T., {et~al.} 2023, \aap, 670, A145

\bibitem[{{Acciari} {et~al.}(2010){Acciari}, {Aliu}, {Arlen}, {Aune},
  {Bautista}, {Beilicke}, {Benbow}, {B{\"o}ttcher}, {Boltuch}, {Bradbury},
  {Buckley}, {Bugaev}, {Byrum}, {Cannon}, {Cesarini}, {Ciupik}, {Cui},
  {Dickherber}, {Duke}, {Falcone}, {Finley}, {Finnegan}, {Fortson}, {Furniss},
  {Galante}, {Gall}, {Gibbs}, {Gillanders}, {Godambe}, {Grube}, {Guenette},
  {Gyuk}, {Hanna}, {Holder}, {Hui}, {Humensky}, {Imran}, {Kaaret}, {Karlsson},
  {Kertzman}, {Kieda}, {Konopelko}, {Krawczynski}, {Krennrich}, {Lang},
  {Lamerato}, {LeBohec}, {Maier}, {McArthur}, {McCann}, {McCutcheon},
  {Moriarty}, {Mukherjee}, {Ong}, {Otte}, {Pandel}, {Perkins}, {Petry},
  {Pichel}, {Pohl}, {Quinn}, {Ragan}, {Reyes}, {Reynolds}, {Roache}, {Rose},
  {Roustazadeh}, {Schroedter}, {Sembroski}, {Senturk}, {Smith}, {Steele},
  {Swordy}, {Te{\v{s}}i{\'c}}, {Theiling}, {Thibadeau}, {Varlotta},
  {Vassiliev}, {Vincent}, {Wagner}, {Wakely}, {Ward}, {Weekes}, {Weinstein},
  {Weisgarber}, {Williams}, {Wissel}, {Wood}, {Zitzer}, {Ackermann}, {Ajello},
  {Antolini}, {Baldini}, {Ballet}, {Barbiellini}, {Bastieri}, {Bechtol},
  {Bellazzini}, {Berenji}, {Blandford}, {Bloom}, {Bonamente}, {Borgland},
  {Bouvier}, {Bregeon}, {Brigida}, {Bruel}, {Buehler}, {Buson}, {Caliandro},
  {Cameron}, {Caraveo}, {Carrigan}, {Casandjian}, {Cavazzuti}, {Cecchi},
  {{\c{C}}elik}, {Charles}, {Chekhtman}, {Cheung}, {Chiang}, {Ciprini},
  {Claus}, {Cohen-Tanugi}, {Conrad}, {Dermer}, {de Palma}, {Silva}, {Drell},
  {Dubois}, {Dumora}, {Farnier}, {Favuzzi}, {Fegan}, {Fortin}, {Frailis},
  {Fukazawa}, {Funk}, {Fusco}, {Gargano}, {Gasparrini}, {Gehrels}, {Germani},
  {Giebels}, {Giglietto}, {Giordano}, {Giroletti}, {Glanzman}, {Godfrey},
  {Grenier}, {Grove}, {Guiriec}, {Hays}, {Horan}, {Hughes}, {J{\'o}hannesson},
  {Johnson}, {Johnson}, {Kamae}, {Katagiri}, {Kataoka}, {Kn{\"o}dlseder},
  {Kuss}, {Lande}, {Latronico}, {Lee}, {Llena Garde}, {Longo}, {Loparco},
  {Lott}, {Lovellette}, {Lubrano}, {Makeev}, {Mazziotta}, {Michelson},
  {Mitthumsiri}, {Mizuno}, {Moiseev}, {Monte}, {Monzani}, {Morselli},
  {Moskalenko}, {Murgia}, {Nolan}, {Norris}, {Nuss}, {Ohno}, {Ohsugi},
  {Omodei}, {Orlando}, {Ormes}, {Paneque}, {Panetta}, {Pelassa}, {Pepe},
  {Pesce-Rollins}, {Piron}, {Porter}, {Rain{\`o}}, \&
  {Rando}}]{2010ApJ...715L..49A}
{Acciari}, V.~A., {Aliu}, E., {Arlen}, T., {et~al.} 2010, \apjl, 715, L49

\bibitem[{{Acciari} {et~al.}(2020){Acciari}, {Ansoldi}, {Antonelli}, {Engels},
  {Asano}, {Baack}, {Babi{\'c}}, {Banerjee}, {Barres de Almeida}, {Barrio},
  {Becerra Gonz{\'a}lez}, {Bednarek}, {Bellizzi}, {Bernardini}, {Berti},
  {Besenrieder}, {Bhattacharyya}, {Bigongiari}, {Biland}, {Blanch}, {Bonnoli},
  {Bo{\v{s}}njak}, {Busetto}, {Carosi}, {Ceribella}, {Cerruti}, {Chai},
  {Chilingaryan}, {Cikota}, {Colak}, {Colin}, {Colombo}, {Contreras},
  {Cortina}, {Covino}, {D'Elia}, {Da Vela}, {Dazzi}, {De Angelis}, {De Lotto},
  {Delfino}, {Delgado}, {Depaoli}, {Di Pierro}, {Di Venere}, {Do Souto
  Espi{\~n}eira}, {Dominis Prester}, {Donini}, {Dorner}, {Doro}, {Elsaesser},
  {Ramazani}, {Fattorini}, {Ferrara}, {Fidalgo}, {Foffano}, {Fonseca}, {Font},
  {Fruck}, {Fukami}, {Garc{\'\i}a L{\'o}pez}, {Garczarczyk}, {Gasparyan},
  {Gaug}, {Giglietto}, {Giordano}, {Godinovi{\'c}}, {Green}, {Guberman},
  {Hadasch}, {Hahn}, {Herrera}, {Hoang}, {Hrupec}, {H{\"u}tten}, {Inada},
  {Inoue}, {Ishio}, {Iwamura}, {Jouvin}, {Kerszberg}, {Kubo}, {Kushida},
  {Lamastra}, {Lelas}, {Leone}, {Lindfors}, {Lombardi}, {Longo}, {L{\'o}pez},
  {L{\'o}pez-Coto}, {L{\'o}pez-Oramas}, {Loporchio}, {Machado de Oliveira
  Fraga}, {Maggio}, {Majumdar}, {Makariev}, {Mallamaci}, {Maneva}, {Manganaro},
  {Mannheim}, {Maraschi}, {Mariotti}, {Mart{\'\i}nez}, {Mazin},
  {Mi{\'c}anovi{\'c}}, {Miceli}, {Minev}, {Miranda}, {Mirzoyan}, {Molina},
  {Moralejo}, {Morcuende}, {Moreno}, {Moretti}, {Munar-Adrover}, {Neustroev},
  {Nigro}, {Nilsson}, {Ninci}, {Nishijima}, {Noda}, {Nogu{\'e}s}, {Nozaki},
  {Paiano}, {Palatiello}, {Paneque}, {Paoletti}, {Paredes}, {Pe{\~n}il},
  {Peresano}, {Persic}, {Prada Moroni}, {Prandini}, {Puljak}, {Rhode},
  {Rib{\'o}}, {Rico}, {Righi}, {Rugliancich}, {Saha}, {Sahakyan}, {Saito},
  {Sakurai}, {Satalecka}, {Schmidt}, {Schweizer}, {Sitarek},
  {{\v{S}}nidari{\'c}}, {Sobczynska}, {Somero}, {Stamerra}, {Strom}, {Strzys},
  {Suda}, {Suri{\'c}}, {Takahashi}, {Tavecchio}, {Temnikov}, {Terzi{\'c}},
  {Teshima}, {Torres-Alb{\`a}}, {Tosti}, {Vagelli}, {van Scherpenberg},
  {Vanzo}, {Vazquez Acosta}, {Vigorito}, {Vitale}, {Vovk}, {Will}, {Zari{\'c}},
  {Arcaro}, {Carosi}, {D'Ammando}, {Tombesi}, \&
  {Lohfink}}]{2020ApJS..247...16A}
{Acciari}, V.~A., {Ansoldi}, S., {Antonelli}, L.~A., {et~al.} 2020, \apjs, 247,
  16

\bibitem[{{Acharyya} {et~al.}(2023){Acharyya}, {Adams}, {Archer}, {Bangale},
  {Bartkoske}, {Batista}, {Benbow}, {Brill}, {Buckley}, {Christiansen},
  {Chromey}, {Errando}, {Falcone}, {Feng}, {Foote}, {Fortson}, {Furniss},
  {Gallagher}, {Hanlon}, {Hanna}, {Hervet}, {Hinrichs}, {Hoang}, {Holder},
  {Humensky}, {Jin}, {Kaaret}, {Kertzman}, {Kherlakian}, {Kieda}, {Kleiner},
  {Korzoun}, {Kumar}, {Lang}, {Lundy}, {Maier}, {McGrath}, {Millard}, {Millis},
  {Mooney}, {Moriarty}, {Mukherjee}, {O'Brien}, {Ong}, {Pohl}, {Pueschel},
  {Quinn}, {Ragan}, {Reynolds}, {Ribeiro}, {Roache}, {Sadeh}, {Sadun}, {Saha},
  {Santander}, {Sembroski}, {Shang}, {Splettstoesser}, {Talluri}, {Tucci},
  {Vassiliev}, {Weinstein}, {Williams}, {Wong}, {Woo}, {Aharonian},
  {Aschersleben}, {Backes}, {Martins}, {Batzofin}, {Becherini}, {Berge},
  {Bernl{\"o}hr}, {Bi}, {B{\"o}ttcher}, {Boisson}, {Bolmont}, {de Bony de
  Lavergne}, {Borowska}, {Bouyahiaoui}, {Bradascio}, {Breuhaus}, {Brose},
  {Brun}, {Bruno}, {Bulik}, {Burger-Scheidlin}, {Caroff}, {Casanova}, {Cecil},
  {Celic}, {Cerruti}, {Chand}, {Chandra}, {Chen}, {Chibueze}, {Chibueze},
  {Cotter}, {Dai}, {Mbarubucyeye}, {Djannati-Ata{\"\i}}, {Dmytriiev},
  {Doroshenko}, {Einecke}, {Ernenwein}, {de Clairfontaine}, {Filipovic},
  {Fontaine}, {F{\"u}{\ss}ling}, {Funk}, {Gabici}, {Ghafourizadeh}, {Giavitto},
  {Glawion}, {Glicenstein}, {Goswami}, {Grolleron}, {Haerer}, {Hinton},
  {Holch}, {Holler}, {Horns}, {Jamrozy}, {Jankowsky}, {Joshi}, {Jung-Richardt},
  {Kasai}, {Katarzy{\'n}ski}, {Khatoon}, {Kh{\'e}lifi}, {Klepser},
  {Klu{\'z}niak}, {Kosack}, {Kostunin}, {Lang}, {Le Stum}, {Lemi{\`e}re},
  {Lenain}, {Leuschner}, {Lohse}, {Luashvili}, {Lypova}, {Mackey}, {Malyshev},
  {Marandon}, {Marchegiani}, {Marcowith}, {Mart{\'\i}-Devesa}, {Marx},
  {Mitchell}, {Moderski}, {Mohrmann}, {Montanari}, {Moulin}, {Murach},
  {Nakashima}, {Niemiec}, {Noel}, {O'Brien}, {Olivera-Nieto}, {de Ona
  Wilhelmi}, {Ostrowski}, {Panny}, {Panter}, {Peron}, {Prokhorov},
  {P{\"u}hlhofer}, {Punch}, {Quirrenbach}, {Reichherzer}, {Reimer}, {Reimer},
  {Ren}, {Renaud}, {Rieger}, {Rudak}, {Ruiz-Velasco}, {Sahakian}, {Santangelo},
  {Sasaki}, {Sch{\"a}fer}, {Sch{\"u}ssler}, {Schutte}, {Schwanke}, {Shapopi},
  {Specovius}, {Spencer}, {Stawarz}, {Steenkamp}, {Steinmassl}, {Sushch},
  {Suzuki}, {Takahashi}, {Tanaka}, {Terrier}, {van Eldik}, {Vecchi}, {Veh},
  {Venter}, \& {Vink}}]{2023ApJ...954...70A}
{Acharyya}, A., {Adams}, C.~B., {Archer}, A., {et~al.} 2023, \apj, 954, 70

\bibitem[{{Ackermann} {et~al.}(2015){Ackermann}, {Ajello}, {Albert}, {Atwood},
  {Baldini}, {Ballet}, {Barbiellini}, {Bastieri}, {Bechtol}, {Bellazzini},
  {Bissaldi}, {Blandford}, {Bloom}, {Bottacini}, {Brandt}, {Bregeon}, {Bruel},
  {Buehler}, {Buson}, {Caliandro}, {Cameron}, {Caragiulo}, {Caraveo},
  {Cavazzuti}, {Cecchi}, {Charles}, {Chekhtman}, {Chiang}, {Chiaro}, {Ciprini},
  {Claus}, {Cohen-Tanugi}, {Conrad}, {Cuoco}, {Cutini}, {D'Ammando}, {de
  Angelis}, {de Palma}, {Dermer}, {Digel}, {Silva}, {Drell}, {Favuzzi},
  {Ferrara}, {Focke}, {Franckowiak}, {Fukazawa}, {Funk}, {Fusco}, {Gargano},
  {Gasparrini}, {Germani}, {Giglietto}, {Giommi}, {Giordano}, {Giroletti},
  {Godfrey}, {Gomez-Vargas}, {Grenier}, {Guiriec}, {Gustafsson}, {Hadasch},
  {Hayashi}, {Hays}, {Hewitt}, {Ippoliti}, {Jogler}, {J{\'o}hannesson},
  {Johnson}, {Johnson}, {Kamae}, {Kataoka}, {Kn{\"o}dlseder}, {Kuss},
  {Larsson}, {Latronico}, {Li}, {Li}, {Longo}, {Loparco}, {Lott}, {Lovellette},
  {Lubrano}, {Madejski}, {Manfreda}, {Massaro}, {Mayer}, {Mazziotta},
  {McEnery}, {Michelson}, {Mitthumsiri}, {Mizuno}, {Moiseev}, {Monzani},
  {Morselli}, {Moskalenko}, {Murgia}, {Nemmen}, {Nuss}, {Ohsugi}, {Omodei},
  {Orlando}, {Ormes}, {Paneque}, {Panetta}, {Perkins}, {Pesce-Rollins},
  {Piron}, {Pivato}, {Porter}, {Rain{\`o}}, {Rando}, {Razzano}, {Razzaque},
  {Reimer}, {Reimer}, {Reposeur}, {Ritz}, {Romani}, {S{\'a}nchez-Conde},
  {Schaal}, {Schulz}, {Sgr{\`o}}, {Siskind}, {Spandre}, {Spinelli}, {Strong},
  {Suson}, {Takahashi}, {Thayer}, {Thayer}, {Tibaldo}, {Tinivella}, {Torres},
  {Tosti}, {Troja}, {Uchiyama}, {Vianello}, {Werner}, {Winer}, {Wood}, {Wood},
  {Zaharijas}, \& {Zimmer}}]{2015ApJ...799...86A}
{Ackermann}, M., {Ajello}, M., {Albert}, A., {et~al.} 2015, \apj, 799, 86

\bibitem[{{Aharonian} {et~al.}(2002){Aharonian}, {Akhperjanian}, {Barrio},
  {Beilicke}, {Bernl{\"o}hr}, {B{\"o}rst}, {Bojahr}, {Bolz}, {Contreras},
  {Cornils}, {Cortina}, {Denninghoff}, {Fonseca}, {Girma}, {Gonzalez},
  {G{\"o}tting}, {Heinzelmann}, {Hermann}, {Heusler}, {Hofmann}, {Horns},
  {Jung}, {Kankanyan}, {Kestel}, {Kettler}, {Kohnle}, {Konopelko}, {Kornmeyer},
  {Kranich}, {Krawczynski}, {Lampeitl}, {Lopez}, {Lorenz}, {Lucarelli},
  {Magnussen}, {Mang}, {Meyer}, {Mirzoyan}, {Moralejo}, {Ona}, {Padilla},
  {Panter}, {Plaga}, {Plyasheshnikov}, {P{\"u}hlhofer}, {Rauterberg},
  {R{\"o}hring}, {Rhode}, {Robrade}, {Rowell}, {Sahakian}, {Samorski},
  {Schilling}, {Schr{\"o}der}, {Sevilla}, {Siems}, {Stamm}, {Tluczykont},
  {V{\"o}lk}, {Wiedner}, \& {Wittek}}]{2002A&A...384L..23A}
{Aharonian}, F., {Akhperjanian}, A., {Barrio}, J., {et~al.} 2002, \aap, 384,
  L23

\bibitem[{{Aharonian} {et~al.}(2010){Aharonian}, {Akhperjanian}, {Anton},
  {Barres de Almeida}, {Bazer-Bachi}, {Becherini}, {Behera}, {Benbow},
  {Bernl{\"o}hr}, {Bochow}, {Boisson}, {Bolmont}, {Borrel}, {Brucker}, {Brun},
  {Brun}, {B{\"u}hler}, {Bulik}, {B{\"u}sching}, {Boutelier}, {Chadwick},
  {Charbonnier}, {Chaves}, {Cheesebrough}, {Chounet}, {Clapson}, {Coignet},
  {Dalton}, {Daniel}, {Davids}, {Degrange}, {Deil}, {Dickinson},
  {Djannati-Ata{\"\i}}, {Domainko}, {O'C. Drury}, {Dubois}, {Dubus}, {Dyks},
  {Dyrda}, {Egberts}, {Emmanoulopoulos}, {Espigat}, {Farnier}, {Feinstein},
  {Fiasson}, {F{\"o}rster}, {Fontaine}, {F{\"u}{\ss}ling}, {Gabici}, {Gallant},
  {G{\'e}rard}, {Gerbig}, {Giebels}, {Glicenstein}, {Gl{\"u}ck}, {Goret},
  {G{\"o}ring}, {Hauser}, {Hauser}, {Heinz}, {Heinzelmann}, {Henri}, {Hermann},
  {Hinton}, {Hoffmann}, {Hofmann}, {Holleran}, {Hoppe}, {Horns},
  {Jacholkowska}, {de Jager}, {Jahn}, {Jung}, {Katarzy{\'n}ski}, {Katz},
  {Kaufmann}, {Kendziorra}, {Kerschhaggl}, {Khangulyan}, {Kh{\'e}lifi},
  {Keogh}, {Klu{\'z}niak}, {Kneiske}, {Komin}, {Kosack}, {Lamanna}, {Lenain},
  {Lohse}, {Marandon}, {Martin}, {Martineau-Huynh}, {Marcowith}, {Masbou},
  {Maurin}, {McComb}, {Medina}, {Moderski}, {Moulin}, {Naumann-Godo}, {de
  Naurois}, {Nedbal}, {Nekrassov}, {Nicholas}, {Niemiec}, {Nolan}, {Ohm},
  {Olive}, {de O{\~n}a Wilhelmi}, {Orford}, {Ostrowski}, {Panter}, {Paz
  Arribas}, {Pedaletti}, {Pelletier}, {Petrucci}, {Pita}, {P{\"u}hlhofer},
  {Punch}, {Quirrenbach}, {Raubenheimer}, {Raue}, {Rayner}, {Renaud}, {Rieger},
  {Ripken}, {Rob}, {Rosier-Lees}, {Rowell}, {Rudak}, {Rulten}, {Ruppel},
  {Sahakian}, {Santangelo}, {Schlickeiser}, {Sch{\"o}ck}, {Schr{\"o}der},
  {Schwanke}, {Schwarzburg}, {Schwemmer}, {Shalchi}, {Sikora}, {Skilton},
  {Sol}, {Spangler}, {Stawarz}, {Steenkamp}, {Stegmann}, {Stinzing},
  {Superina}, {Szostek}, {Tam}, {Tavernet}, {Terrier}, {Tibolla}, {Tluczykont},
  {van Eldik}, {Vasileiadis}, {Venter}, {Venter}, {Vialle}, {Vincent},
  {Vivier}, {V{\"o}lk}, {Volpe}, {Wagner}, {Ward}, {Zdziarski}, \&
  {Zech}}]{2010A&A...521A..69A}
{Aharonian}, F., {Akhperjanian}, A.~G., {Anton}, G., {et~al.} 2010, \aap, 521,
  A69

\bibitem[{{Aharonian} {et~al.}(2006){Aharonian}, {Akhperjanian}, {Bazer-Bachi},
  {Beilicke}, {Benbow}, {Berge}, {Bernl{\"o}hr}, {Boisson}, {Bolz}, {Borrel},
  {Braun}, {Breitling}, {Brown}, {Chadwick}, {Chounet}, {Cornils},
  {Costamante}, {Degrange}, {Dickinson}, {Djannati-Ata{\"\i}}, {Drury},
  {Dubus}, {Emmanoulopoulos}, {Espigat}, {Feinstein}, {Fontaine}, {Fuchs},
  {Funk}, {Gallant}, {Giebels}, {Gillessen}, {Glicenstein}, {Goret},
  {Hadjichristidis}, {Hauser}, {Hauser}, {Heinzelmann}, {Henri}, {Hermann},
  {Hinton}, {Hofmann}, {Holleran}, {Horns}, {Jacholkowska}, {de Jager},
  {Kh{\'e}lifi}, {Klages}, {Komin}, {Konopelko}, {Latham}, {Le Gallou},
  {Lemi{\`e}re}, {Lemoine-Goumard}, {Leroy}, {Lohse}, {Martin},
  {Martineau-Huynh}, {Marcowith}, {Masterson}, {McComb}, {de Naurois}, {Nolan},
  {Noutsos}, {Orford}, {Osborne}, {Ouchrif}, {Panter}, {Pelletier}, {Pita},
  {P{\"u}hlhofer}, {Punch}, {Raubenheimer}, {Raue}, {Raux}, {Rayner}, {Reimer},
  {Reimer}, {Ripken}, {Rob}, {Rolland}, {Rowell}, {Sahakian}, {Saug{\'e}},
  {Schlenker}, {Schlickeiser}, {Schuster}, {Schwanke}, {Siewert}, {Sol},
  {Spangler}, {Steenkamp}, {Stegmann}, {Tavernet}, {Terrier}, {Th{\'e}oret},
  {Tluczykont}, {van Eldik}, {Vasileiadis}, {Venter}, {Vincent}, {V{\"o}lk}, \&
  {Wagner}}]{2006Natur.440.1018A}
{Aharonian}, F., {Akhperjanian}, A.~G., {Bazer-Bachi}, A.~R., {et~al.} 2006,
  \nat, 440, 1018

\bibitem[{{Aharonian} {et~al.}(2023){Aharonian}, {Aschersleben}, {Backes},
  {Martins}, {Batzofin}, {Becherini}, {Berge}, {Bi}, {Bouyahiaoui}, {Breuhaus},
  {Brose}, {Brun}, {Bruno}, {Bulik}, {Burger-Scheidlin}, {Bylund}, {Caroff},
  {Casanova}, {Celic}, {Cerruti}, {Chand}, {Chandra}, {Chen}, {Chibueze},
  {Chibueze}, {Cotter}, {de Bony}, {Egberts}, {Ernenwein}, {Fichet de
  Clairfontaine}, {Filipovic}, {Fontaine}, {F{\"u}ssling}, {Funk}, {Gabici},
  {Ghafourizadeh}, {Giavitto}, {Glawion}, {Glicenstein}, {Goswami}, {Grondin},
  {Haerer}, {Holch}, {Holler}, {Horns}, {Jamrozy}, {Jankowsky}, {Joshi},
  {Jung-Richardt}, {Kasai}, {Katarzy{\'n}ski}, {Khatoon}, {Kh{\'e}lifi},
  {Klu{\'z}niak}, {Komin}, {Kosack}, {Kostunin}, {Lang}, {Le Stum}, {Leitl},
  {Lemi{\`e}re}, {Lenain}, {Leuschner}, {Lohse}, {Luashvili}, {Lypova},
  {Mackey}, {Malyshev}, {Malyshev}, {Marandon}, {Marchegiani}, {Marcowith},
  {Mart{\'\i}-Devesa}, {Marx}, {Meyer}, {Mitchell}, {Moderski}, {Mohrmann},
  {Montanari}, {Moulin}, {Muller}, {Murach}, {Nakashima}, {Niemiec}, {Ohm},
  {Olivera-Nieto}, {de Ona Wilhelmi}, {Panny}, {Panter}, {Parsons}, {Peron},
  {Prokhorov}, {Prokoph}, {P{\"u}hlhofer}, {Punch}, {Quirrenbach},
  {Reichherzer}, {Reimer}, {Reimer}, {Reville}, {Rieger}, {Rowell}, {Rudak},
  {Ruiz-Velasco}, {Sahakian}, {Sanchez}, {Sasaki}, {Sch{\"u}ssler}, {Schutte},
  {Schwanke}, {Shapopi}, {Sol}, {Spencer}, {Steinmassl}, {Suzuki}, {Takahashi},
  {Tanaka}, {Taylor}, {Terrier}, {Thorpe-Morgan}, {Tsirou}, {Tsuji},
  {Uchiyama}, {van Eldik}, {Veh}, {Venter}, {Wagner}, {White}, {Wierzcholska},
  {Wong}, {Zacharias}, {Zargaryan}, {Zdziarski}, {Zouari}, {{\.Z}ywucka},
  {Meyer}, \& {Fermi-LAT Collaboration}}]{2023ApJ...950L..16A}
{Aharonian}, F., {Aschersleben}, J., {Backes}, M., {et~al.} 2023, \apjl, 950,
  L16

\bibitem[{{Ajello} {et~al.}(2015){Ajello}, {Gasparrini}, {S{\'a}nchez-Conde},
  {Zaharijas}, {Gustafsson}, {Cohen-Tanugi}, {Dermer}, {Inoue}, {Hartmann},
  {Ackermann}, {Bechtol}, {Franckowiak}, {Reimer}, {Romani}, \&
  {Strong}}]{2015ApJ...800L..27A}
{Ajello}, M., {Gasparrini}, D., {S{\'a}nchez-Conde}, M., {et~al.} 2015, \apjl,
  800, L27

\bibitem[{{Albareti} {et~al.}(2017){Albareti}, {Allende Prieto}, {Almeida},
  {Anders}, {Anderson}, {Andrews}, {Arag{\'o}n-Salamanca},
  {Argudo-Fern{\'a}ndez}, {Armengaud}, {Aubourg}, {Avila-Reese}, {Badenes},
  {Bailey}, {Barbuy}, {Barger}, {Barrera-Ballesteros}, {Bartosz}, {Basu},
  {Bates}, {Battaglia}, {Baumgarten}, {Baur}, {Bautista}, {Beers}, {Belfiore},
  {Bershady}, {Bertran de Lis}, {Bird}, {Bizyaev}, {Blanc}, {Blanton},
  {Blomqvist}, {Bolton}, {Borissova}, {Bovy}, {Brandt}, {Brinkmann},
  {Brownstein}, {Bundy}, {Burtin}, {Busca}, {Camacho Chavez}, {Cano D{\'\i}az},
  {Cappellari}, {Carrera}, {Chen}, {Cherinka}, {Cheung}, {Chiappini},
  {Chojnowski}, {Chuang}, {Chung}, {Cirolini}, {Clerc}, {Cohen}, {Comerford},
  {Comparat}, {Correa do Nascimento}, {Cousinou}, {Covey}, {Crane}, {Croft},
  {Cunha}, {Darling}, {Davidson}, {Dawson}, {Da Costa}, {Da Silva Ilha},
  {Deconto Machado}, {Delubac}, {De Lee}, {De la Macorra}, {De la Torre},
  {Diamond-Stanic}, {Donor}, {Downes}, {Drory}, {Du}, {Du Mas des Bourboux},
  {Dwelly}, {Ebelke}, {Eigenbrot}, {Eisenstein}, {Elsworth}, {Emsellem},
  {Eracleous}, {Escoffier}, {Evans}, {Falc{\'o}n-Barroso}, {Fan}, {Favole},
  {Fernandez-Alvar}, {Fernandez-Trincado}, {Feuillet}, {Fleming},
  {Font-Ribera}, {Freischlad}, {Frinchaboy}, {Fu}, {Gao}, {Garcia},
  {Garcia-Dias}, {Garcia-Hern{\'a}ndez}, {Garcia P{\'e}rez}, {Gaulme}, {Ge},
  {Geisler}, {Gillespie}, {Gil Marin}, {Girardi}, {Goddard}, {Gomez Maqueo
  Chew}, {Gonzalez-Perez}, {Grabowski}, {Green}, {Grier}, {Grier}, {Guo},
  {Guy}, {Hagen}, {Hall}, {Harding}, {Harley}, {Hasselquist}, {Hawley},
  {Hayes}, {Hearty}, {Hekker}, {Hernandez Toledo}, {Ho}, {Hogg},
  {Holley-Bockelmann}, {Holtzman}, {Holzer}, {Hu}, {Huber}, {Hutchinson},
  {Hwang}, {Ibarra-Medel}, {Ivans}, {Ivory}, {Jaehnig}, {Jensen}, {Johnson},
  {Jones}, {Jullo}, {Kallinger}, {Kinemuchi}, {Kirkby}, {Klaene}, {Kneib},
  {Kollmeier}, {Lacerna}, {Lane}, {Lang}, {Laurent}, {Law}, {Leauthaud}, {Le
  Goff}, {Li}, {Li}, {Li}, {Li}, {Liang}, {Liang}, {Lima}, {Lin}, {Lin}, {Lin},
  {Liu}, {Long}, {Lucatello}, {MacDonald}, {MacLeod}, {Mackereth}, {Mahadevan},
  {Maia}, {Maiolino}, {Majewski}, {Malanushenko}, {Malanushenko}, {Mallmann},
  {Manchado}, {Maraston}, {Marques-Chaves}, {Martinez Valpuesta}, {Masters},
  {Mathur}, {McGreer}, {Merloni}, {Merrifield}, {M{\'e}sz{\'a}ros}, {Meza},
  {Miglio}, {Minchev}, {Molaverdikhani}, {Montero-Dorta}, {Mosser}, {Muna}, \&
  {Myers}}]{2017ApJS..233...25A}
{Albareti}, F.~D., {Allende Prieto}, C., {Almeida}, A., {et~al.} 2017, \apjs,
  233, 25

\bibitem[{{Albert} {et~al.}(2021){Albert}, {Alvarez}, {Angeles Camacho},
  {Arteaga-Vel{\'a}zquez}, {Arunbabu}, {Avila Rojas}, {Ayala Solares},
  {Baghmanyan}, {Belmont-Moreno}, {BenZvi}, {Brisbois}, {Caballero-Mora},
  {Capistr{\'a}n}, {Carrami{\~n}ana}, {Casanova}, {Cotti}, {Cotzomi},
  {Couti{\~n}o de Le{\'o}n}, {De la Fuente}, {Dingus}, {DuVernois}, {Durocher},
  {D{\'\i}az-V{\'e}lez}, {Engel}, {Espinoza}, {Fan}, {Fern{\'a}ndez Alonso},
  {Fleischhack}, {Fraija}, {Galv{\'a}n-G{\'a}mez}, {Garc{\'\i}a},
  {Garc{\'\i}a-Gonz{\'a}lez}, {Garfias}, {Gonz{\'a}lez}, {Goodman}, {Harding},
  {Hern{\'a}ndez}, {Hona}, {Huang}, {Hueyotl-Zahuantitla}, {H{\"u}ntemeyer},
  {Iriarte}, {Jardin-Blicq}, {Joshi}, {Kieda}, {Kunde}, {Lara}, {Lee},
  {Le{\'o}n Vargas}, {Linnemann}, {Longinotti}, {Luis-Raya}, {Lundeen},
  {Malone}, {Mart{\'\i}nez}, {Martinez-Castellanos}, {Mart{\'\i}nez-Castro},
  {Matthews}, {Miranda-Romagnoli}, {Morales-Soto}, {Moreno}, {Mostaf{\'a}},
  {Nayerhoda}, {Nellen}, {Newbold}, {Nisa}, {Noriega-Papaqui}, {Olivera-Nieto},
  {Peisker}, {P{\'e}rez-P{\'e}rez}, {Rho}, {Rosa-Gonz{\'a}lez}, {Ruiz-Velasco},
  {Salazar}, {Greus}, {Sandoval}, {Schneider}, {Schoorlemmer}, {Smith},
  {Springer}, {Tollefson}, {Torres}, {Torres-Escobedo}, {Ure{\~n}a-Mena},
  {Villase{\~n}or}, {Weisgarber}, {Willox}, {Zepeda}, {Zhou}, {de Le{\'o}n}, \&
  {HAWC Collaboration}}]{2021ApJ...907...67A}
{Albert}, A., {Alvarez}, C., {Angeles Camacho}, J.~R., {et~al.} 2021, \apj,
  907, 67

\bibitem[{{Atwood} {et~al.}(2009){Atwood}, {Abdo}, {Ackermann}, {Althouse},
  {Anderson}, {Axelsson}, {Baldini}, {Ballet}, {Band}, {Barbiellini},
  {Bartelt}, {Bastieri}, {Baughman}, {Bechtol}, {B{\'e}d{\'e}r{\`e}de},
  {Bellardi}, {Bellazzini}, {Berenji}, {Bignami}, {Bisello}, {Bissaldi},
  {Blandford}, {Bloom}, {Bogart}, {Bonamente}, {Bonnell}, {Borgland},
  {Bouvier}, {Bregeon}, {Brez}, {Brigida}, {Bruel}, {Burnett}, {Busetto},
  {Caliandro}, {Cameron}, {Caraveo}, {Carius}, {Carlson}, {Casandjian},
  {Cavazzuti}, {Ceccanti}, {Cecchi}, {Charles}, {Chekhtman}, {Cheung},
  {Chiang}, {Chipaux}, {Cillis}, {Ciprini}, {Claus}, {Cohen-Tanugi},
  {Condamoor}, {Conrad}, {Corbet}, {Corucci}, {Costamante}, {Cutini}, {Davis},
  {Decotigny}, {DeKlotz}, {Dermer}, {de Angelis}, {Digel}, {do Couto e Silva},
  {Drell}, {Dubois}, {Dumora}, {Edmonds}, {Fabiani}, {Farnier}, {Favuzzi},
  {Flath}, {Fleury}, {Focke}, {Funk}, {Fusco}, {Gargano}, {Gasparrini},
  {Gehrels}, {Gentit}, {Germani}, {Giebels}, {Giglietto}, {Giommi}, {Giordano},
  {Glanzman}, {Godfrey}, {Grenier}, {Grondin}, {Grove}, {Guillemot}, {Guiriec},
  {Haller}, {Harding}, {Hart}, {Hays}, {Healey}, {Hirayama}, {Hjalmarsdotter},
  {Horn}, {Hughes}, {J{\'o}hannesson}, {Johansson}, {Johnson}, {Johnson},
  {Johnson}, {Johnson}, {Kamae}, {Katagiri}, {Kataoka}, {Kavelaars}, {Kawai},
  {Kelly}, {Kerr}, {Klamra}, {Kn{\"o}dlseder}, {Kocian}, {Komin}, {Kuehn},
  {Kuss}, {Landriu}, {Latronico}, {Lee}, {Lee}, {Lemoine-Goumard}, {Lionetto},
  {Longo}, {Loparco}, {Lott}, {Lovellette}, {Lubrano}, {Madejski}, {Makeev},
  {Marangelli}, {Massai}, {Mazziotta}, {McEnery}, {Menon}, {Meurer},
  {Michelson}, {Minuti}, {Mirizzi}, {Mitthumsiri}, {Mizuno}, {Moiseev},
  {Monte}, {Monzani}, {Moretti}, {Morselli}, {Moskalenko}, {Murgia},
  {Nakamori}, {Nishino}, {Nolan}, {Norris}, {Nuss}, {Ohno}, {Ohsugi}, {Omodei},
  {Orlando}, {Ormes}, {Paccagnella}, {Paneque}, {Panetta}, {Parent}, {Pearce},
  {Pepe}, {Perazzo}, {Pesce-Rollins}, {Picozza}, {Pieri}, {Pinchera}, {Piron},
  {Porter}, {Poupard}, {Rain{\`o}}, {Rando}, {Rapposelli}, {Razzano}, {Reimer},
  {Reimer}, {Reposeur}, {Reyes}, {Ritz}, {Rochester}, {Rodriguez}, {Romani},
  {Roth}, {Russell}, {Ryde}, {Sabatini}, {Sadrozinski}, {Sanchez}, {Sander},
  {Sapozhnikov}, {Parkinson}, {Scargle}, {Schalk}, \&
  {Scolieri}}]{2009ApJ...697.1071A}
{Atwood}, W.~B., {Abdo}, A.~A., {Ackermann}, M., {et~al.} 2009, \apj, 697, 1071

\bibitem[{Ballet {et~al.}(2023)Ballet, Bruel, Burnett, \&
  Lott}]{Ballet:2023qzs}
Ballet, J., Bruel, P., Burnett, T.~H., \& Lott, B. 2023
  [\eprint[arXiv]{2307.12546}]

\bibitem[{{Benbow}(2011)}]{2011ICRC....8...51B}
{Benbow}, W. 2011, in International Cosmic Ray Conference, Vol.~8,
  International Cosmic Ray Conference, 51

\bibitem[{{Cao} {et~al.}(2024){Cao}, {Aharonian}, {An}, {Axikegu}, {Bai},
  {Bao}, {Bastieri}, {Bi}, {Bi}, {Cai}, {Cao}, {Cao}, {Cao}, {Chang}, {Chang},
  {Chen}, {Chen}, {Chen}, {Chen}, {Chen}, {Chen}, {Chen}, {Chen}, {Chen},
  {Chen}, {Chen}, {Chen}, {Cheng}, {Cheng}, {Cui}, {Cui}, {Cui}, {Cui}, {Dai},
  {Dai}, {Dai}, {Danzengluobu}, {Della Volpe}, {Dong}, {Duan}, {Fan}, {Fan},
  {Fang}, {Fang}, {Feng}, {Feng}, {Feng}, {Feng}, {Feng}, {Gabici}, {Gao},
  {Gao}, {Gao}, {Gao}, {Gao}, {Gao}, {Ge}, {Geng}, {Giacinti}, {Gong}, {Gou},
  {Gu}, {Guo}, {Guo}, {Guo}, {Guo}, {Han}, {He}, {He}, {He}, {He}, {He},
  {Heller}, {Hor}, {Hou}, {Hou}, {Hou}, {Hu}, {Hu}, {Hu}, {Huang}, {Huang},
  {Huang}, {Huang}, {Huang}, {Huang}, {Huang}, {Ji}, {Jia}, {Jia}, {Jiang},
  {Jiang}, {Jiang}, {Jin}, {Kang}, {Ke}, {Kuleshov}, {Kurinov}, {Li}, {Li},
  {Li}, {Li}, {Li}, {Li}, {Li}, {Li}, {Li}, {Li}, {Li}, {Li}, {Li}, {Li}, {Li},
  {Li}, {Li}, {Li}, {Li}, {Liang}, {Liang}, {Lin}, {Liu}, {Liu}, {Liu}, {Liu},
  {Liu}, {Liu}, {Liu}, {Liu}, {Liu}, {Liu}, {Liu}, {Liu}, {Liu}, {Liu}, {Lu},
  {Luo}, {Lv}, {Ma}, {Ma}, {Ma}, {Mao}, {Min}, {Mitthumsiri}, {Mu}, {Nan},
  {Neronov}, {Ou}, {Pang}, {Pattarakijwanich}, {Pei}, {Qi}, {Qi}, {Qiao},
  {Qin}, {Ruffolo}, {S{\'a}iz}, {Semikoz}, {Shao}, {Shao}, {Shchegolev},
  {Sheng}, {Shu}, {Song}, {Stenkin}, {Stepanov}, {Su}, {Sun}, {Sun}, {Sun},
  {Tam}, {Tang}, {Tang}, {Tian}, {Wang}, {Wang}, {Wang}, {Wang}, {Wang},
  {Wang}, {Wang}, {Wang}, {Wang}, {Wang}, {Wang}, {Wang}, {Wang}, {Wang},
  {Wang}, {Wang}, {Wang}, {Wang}, {Wang}, {Wang}, {Wang}, {Wei}, {Wei}, {Wei},
  {Wen}, {Wu}, \& {Wu}}]{2024ApJS..271...25C}
{Cao}, Z., {Aharonian}, F., {An}, Q., {et~al.} 2024, \apjs, 271, 25

\bibitem[{{Chang} {et~al.}(2019){Chang}, {Arsioli}, {Giommi}, {Padovani}, \&
  {Brandt}}]{2019A&A...632A..77C}
{Chang}, Y.~L., {Arsioli}, B., {Giommi}, P., {Padovani}, P., \& {Brandt}, C.~H.
  2019, \aap, 632, A77

\bibitem[{{Cherenkov Telescope Array Consortium} {et~al.}(2019){Cherenkov
  Telescope Array Consortium}, {Acharya}, {Agudo}, {Al Samarai}, {Alfaro},
  {Alfaro}, {Alispach}, {Alves Batista}, {Amans}, {Amato}, {Ambrosi},
  {Antolini}, {Antonelli}, {Aramo}, {Araya}, {Armstrong}, {Arqueros},
  {Arrabito}, {Asano}, {Ashley}, {Backes}, {Balazs}, {Balbo}, {Ballester},
  {Ballet}, {Bamba}, {Barkov}, {Barres de Almeida}, {Barrio}, {Bastieri},
  {Becherini}, {Belfiore}, {Benbow}, {Berge}, {Bernardini}, {Bernardini},
  {Bernardos}, {Bernl{\"o}hr}, {Bertucci}, {Biasuzzi}, {Bigongiari}, {Biland},
  {Bissaldi}, {Biteau}, {Blanch}, {Blazek}, {Boisson}, {Bolmont}, {Bonanno},
  {Bonardi}, {Bonavolont{\`a}}, {Bonnoli}, {Bosnjak}, {B{\"o}ttcher},
  {Braiding}, {Bregeon}, {Brill}, {Brown}, {Brun}, {Brunetti}, {Buanes},
  {Buckley}, {Bugaev}, {B{\"u}hler}, {Bulgarelli}, {Bulik}, {Burton},
  {Burtovoi}, {Busetto}, {Canestrari}, {Capalbi}, {Capitanio}, {Caproni},
  {Caraveo}, {C{\'a}rdenas}, {Carlile}, {Carosi}, {Carqu{\'\i}n}, {Carr},
  {Casanova}, {Cascone}, {Catalani}, {Catalano}, {Cauz}, {Cerruti}, {Chadwick},
  {Chaty}, {Chaves}, {Chen}, {Chen}, {Chernyakova}, {Chikawa}, {Christov},
  {Chudoba}, {Cie{\'s}lar}, {Coco}, {Colafrancesco}, {Colin}, {Conforti},
  {Connaughton}, {Conrad}, {Contreras}, {Cortina}, {Costa}, {Costantini},
  {Cotter}, {Covino}, {Crocker}, {Cuadra}, {Cuevas}, {Cumani}, {D'A{\`\i}},
  {D'Ammando}, {D'Avanzo}, {D'Urso}, {Daniel}, {Davids}, {Dawson}, {Dazzi}, {De
  Angelis}, {de C{\'a}ssia dos Anjos}, {De Cesare}, {De Franco}, {de Gouveia
  Dal Pino}, {de la Calle}, {de los Reyes Lopez}, {De Lotto}, {De Luca}, {De
  Lucia}, {de Naurois}, {de O{\~n}a Wilhelmi}, {De Palma}, {De Persio}, {de
  Souza}, {Deil}, {Del Santo}, {Delgado}, {della Volpe}, {Di Girolamo}, {Di
  Pierro}, {Di Venere}, {D{\'\i}az}, {Dib}, {Diebold}, {Djannati-Ata{\"\i}},
  {Dom{\'\i}nguez}, {Dominis Prester}, {Dorner}, {Doro}, {Drass}, {Dravins},
  {Dubus}, {Dwarkadas}, {Ebr}, {Eckner}, {Egberts}, {Einecke}, {Ekoume},
  {Els{\"a}sser}, {Ernenwein}, {Espinoza}, {Evoli}, {Fairbairn},
  {Falceta-Goncalves}, {Falcone}, {Farnier}, {Fasola}, {Fedorova}, {Fegan},
  {Fernandez-Alonso}, {Fern{\'a}ndez-Barral}, {Ferrand}, {Fesquet},
  {Filipovic}, {Fioretti}, {Fontaine}, {Fornasa}, {Fortson}, {Freixas
  Coromina}, {Fruck}, {Fujita}, {Fukazawa}, {Funk}, {F{\"u}{\ss}ling},
  {Gabici}, {Gadola}, {Gallant}, {Garcia}, {Garcia L{\'o}pez}, {Garczarczyk},
  {Gaskins}, {Gasparetto}, {Gaug}, {Gerard}, {Giavitto}, {Giglietto}, {Giommi},
  {Giordano}, {Giro}, \& {Giroletti}}]{2019scta.book.....C}
{Cherenkov Telescope Array Consortium}, {Acharya}, B.~S., {Agudo}, I., {et~al.}
  2019, {Science with the Cherenkov Telescope Array}

\bibitem[{{Cortina} \& {CTAO LST Collaboration}(2023)}]{2023ATel16381....1C}
{Cortina}, J. \& {CTAO LST Collaboration}. 2023, The Astronomer's Telegram,
  16381, 1

\bibitem[{Dom\'\i{}nguez {et~al.}(2023)}]{Dominguez:2023rxa}
Dom\'\i{}nguez, A. {et~al.} 2023, Mon. Not. Roy. Astron. Soc., 527, 4632

\bibitem[{{Fossati} {et~al.}(1998){Fossati}, {Maraschi}, {Celotti}, {Comastri},
  \& {Ghisellini}}]{1998MNRAS.299..433F}
{Fossati}, G., {Maraschi}, L., {Celotti}, A., {Comastri}, A., \& {Ghisellini},
  G. 1998, \mnras, 299, 433

\bibitem[{{Franceschini} \& {Rodighiero}(2017)}]{2017A&A...603A..34F}
{Franceschini}, A. \& {Rodighiero}, G. 2017, \aap, 603, A34

\bibitem[{{Furniss} {et~al.}(2019){Furniss}, {Worseck}, {Fumagalli}, {Johnson},
  {Williams}, {Pontrelli}, \& {Prochaska}}]{2019AJ....157...41F}
{Furniss}, A., {Worseck}, G., {Fumagalli}, M., {et~al.} 2019, \aj, 157, 41

\bibitem[{{H.~E.~S.~S. Collaboration} {et~al.}(2025){H.~E.~S.~S.
  Collaboration}, {Aharonian}, {Ait Benkhali}, {Aschersleben}, {Ashkar},
  {Backes}, {Barbosa Martins}, {Batzofin}, {Becherini}, {Berge},
  {Bernl{\"o}hr}, {B{\"o}ttcher}, {Boisson}, {Bolmont}, {de Bony de Lavergne},
  {Borowska}, {Bouyahiaoui}, {Bradascio}, {Brose}, {Brown}, {Brun}, {Bruno},
  {Bulik}, {Burger-Scheidlin}, {Bylund}, {Casanova}, {Celic}, {Cerruti},
  {Chand}, {Chandra}, {Chen}, {Chibueze}, {Chibueze}, {Cotter}, {Cristofari},
  {Mbarubucyeye}, {Davids}, {Devin}, {Djuvsland}, {Dmytriiev}, {Egberts},
  {Einecke}, {Fegan}, {Fontaine}, {Funk}, {Gabici}, {Glicenstein}, {Glombitza},
  {Goswami}, {Grolleron}, {Haerer}, {He{\ss}}, {Hinton}, {Hofmann}, {Holch},
  {Holler}, {Horns}, {Huang}, {Jamrozy}, {Jankowsky}, {Jardin-Blicq}, {Kasai},
  {Katarzy{\'n}ski}, {Khatoon}, {Kh{\'e}lifi}, {Komin}, {Kosack}, {Kostunin},
  {Kundu}, {Lang}, {Le Stum}, {Lemi{\`e}re}, {Lemoine-Goumard}, {Lenain},
  {Leuschner}, {Luashvili}, {Mackey}, {Malyshev}, {Marandon},
  {Mart{\'\i}-Devesa}, {Marx}, {Mehta}, {Mitchell}, {Moderski}, {Mohrmann},
  {Montanari}, {de Naurois}, {Niemiec}, {O'Brien}, {Olivera-Nieto}, {de Ona
  Wilhelmi}, {Ostrowski}, {Panny}, {Panter}, {Pensec}, {P{\"u}hlhofer},
  {Punch}, {Quirrenbach}, {Ravikularaman}, {Regeard}, {Reimer}, {Reimer},
  {Reis}, {Ren}, {Reville}, {Rieger}, {Rowell}, {Rudak}, {Ruiz-Velasco},
  {Sahakian}, {Salzmann}, {Sanchez}, {Santangelo}, {Sasaki}, {Sch{\"a}fer},
  {Sch{\"u}ssler}, {Shapopi}, {Sharma}, {Sol}, {Spencer}, {Stawarz},
  {Steenkamp}, {Steinmassl}, {Steppa}, {Suzuki}, {Takahashi}, {Tanaka},
  {Taylor}, {Terrier}, {Thakur}, {Tsirou}, {van Eldik}, {Vecchi}, {Venter},
  {Vink}, {V{\"o}lk}, {Wach}, {Wagner}, {Wierzcholska}, {Zacharias},
  {Zdziarski}, {Zech}, \& {{\.Z}ywucka}}]{2025A&A...695A.261H}
{H.~E.~S.~S. Collaboration}, {Aharonian}, F., {Ait Benkhali}, F., {et~al.}
  2025, \aap, 695, A261

\bibitem[{{Huentemeyer} {et~al.}(2019){Huentemeyer}, {BenZvi}, {Dingus},
  {Fleischhack}, {Schoorlemmer}, \& {Weisgarber}}]{2019BAAS...51g.109H}
{Huentemeyer}, P., {BenZvi}, S., {Dingus}, B., {et~al.} 2019, in Bulletin of
  the American Astronomical Society, Vol.~51, 109

\bibitem[{{IceCube Collaboration} {et~al.}(2018){IceCube Collaboration},
  {Aartsen}, {Ackermann}, {Adams}, {Aguilar}, {Ahlers}, {Ahrens}, {Al Samarai},
  {Altmann}, {Andeen}, {Anderson}, {Ansseau}, {Anton}, {Arg{\"u}elles},
  {Auffenberg}, {Axani}, {Bagherpour}, {Bai}, {Barron}, {Barwick}, {Baum},
  {Bay}, {Beatty}, {Becker Tjus}, {Becker}, {BenZvi}, {Berley}, {Bernardini},
  {Besson}, {Binder}, {Bindig}, {Blaufuss}, {Blot}, {Bohm}, {B{\"o}rner},
  {Bos}, {B{\"o}ser}, {Botner}, {Bourbeau}, {Bourbeau}, {Bradascio}, {Braun},
  {Brenzke}, {Bretz}, {Bron}, {Brostean-Kaiser}, {Burgman}, {Busse}, {Carver},
  {Cheung}, {Chirkin}, {Christov}, {Clark}, {Classen}, {Coenders}, {Collin},
  {Conrad}, {Coppin}, {Correa}, {Cowen}, {Cross}, {Dave}, {Day}, {de
  Andr{\'e}}, {De Clercq}, {DeLaunay}, {Dembinski}, {De Ridder}, {Desiati}, {de
  Vries}, {de Wasseige}, {de With}, {DeYoung}, {D{\'\i}az-V{\'e}lez}, {di
  Lorenzo}, {Dujmovic}, {Dumm}, {Dunkman}, {Dvorak}, {Eberhardt}, {Ehrhardt},
  {Eichmann}, {Eller}, {Evenson}, {Fahey}, {Fazely}, {Felde}, {Filimonov},
  {Finley}, {Flis}, {Franckowiak}, {Friedman}, {Fritz}, {Gaisser}, {Gallagher},
  {Gerhardt}, {Ghorbani}, {Glauch}, {Gl{\"u}senkamp}, {Goldschmidt},
  {Gonzalez}, {Grant}, {Griffith}, {Haack}, {Hallgren}, {Halzen}, {Hanson},
  {Hebecker}, {Heereman}, {Helbing}, {Hellauer}, {Hickford}, {Hignight},
  {Hill}, {Hoffman}, {Hoffmann}, {Hoinka}, {Hokanson-Fasig}, {Hoshina},
  {Huang}, {Huber}, {Hultqvist}, {H{\"u}nnefeld}, {Hussain}, {In}, {Iovine},
  {Ishihara}, {Jacobi}, {Japaridze}, {Jeong}, {Jero}, {Jones}, {Kalaczynski},
  {Kang}, {Kappes}, {Kappesser}, {Karg}, {Karle}, {Katz}, {Kauer}, {Keivani},
  {Kelley}, {Kheirandish}, {Kim}, {Kim}, {Kintscher}, {Kiryluk}, {Kittler},
  {Klein}, {Koirala}, {Kolanoski}, {K{\"o}pke}, {Kopper}, {Kopper},
  {Koschinsky}, {Koskinen}, {Kowalski}, {Krings}, {Kroll}, {Kr{\"u}ckl},
  {Kunwar}, {Kurahashi}, {Kuwabara}, {Kyriacou}, {Labare}, {Lanfranchi},
  {Larson}, {Lauber}, {Leonard}, {Lesiak-Bzdak}, {Leuermann}, {Liu}, {Lozano
  Mariscal}, {Lu}, {L{\"u}nemann}, {Luszczak}, {Madsen}, {Maggi}, {Mahn},
  {Mancina}, {Maruyama}, {Mase}, {Maunu}, {Meagher}, {Medici}, {Meier},
  {Menne}, {Merino}, {Meures}, {Miarecki}, {Micallef}, {Moment{\'e}},
  {Montaruli}, {Moore}, {Morse}, {Moulai}, {Nahnhauer}, {Nakarmi}, {Naumann},
  \& {Neer}}]{2018Sci...361.1378I}
{IceCube Collaboration}, {Aartsen}, M.~G., {Ackermann}, M., {et~al.} 2018,
  Science, 361, eaat1378

\bibitem[{{Kerby} {et~al.}(2021){Kerby}, {Kaur}, {Falcone}, {Eskenasy},
  {Hancock}, {Stroh}, {Ferrara}, {Ray}, {Kennea}, \&
  {Grove}}]{2021ApJ...923...75K}
{Kerby}, S., {Kaur}, A., {Falcone}, A.~D., {et~al.} 2021, \apj, 923, 75

\bibitem[{{Lefaucheur} \& {Pita}(2017)}]{2017A&A...602A..86L}
{Lefaucheur}, J. \& {Pita}, S. 2017, \aap, 602, A86

\bibitem[{{Lyke} {et~al.}(2020){Lyke}, {Higley}, {McLane}, {Schurhammer},
  {Myers}, {Ross}, {Dawson}, {Chabanier}, {Martini}, {Busca}, {Mas des
  Bourboux}, {Salvato}, {Streblyanska}, {Zarrouk}, {Burtin}, {Anderson},
  {Bautista}, {Bizyaev}, {Brandt}, {Brinkmann}, {Brownstein}, {Comparat},
  {Green}, {de la Macorra}, {Mu{\~n}oz Guti{\'e}rrez}, {Hou}, {Newman},
  {Palanque-Delabrouille}, {P{\^a}ris}, {Percival}, {Petitjean}, {Rich},
  {Rossi}, {Schneider}, {Smith}, {Vivek}, \& {Weaver}}]{2020ApJS..250....8L}
{Lyke}, B.~W., {Higley}, A.~N., {McLane}, J.~N., {et~al.} 2020, \apjs, 250, 8

\bibitem[{{Mao}(2011)}]{2011NewA...16..503M}
{Mao}, L.~S. 2011, \na, 16, 503

\bibitem[{{Mart{\'\i}nez-Huerta} {et~al.}(2019){Mart{\'\i}nez-Huerta},
  {Biteau}, {Lefaucheur}, {Meyer}, {Pita}, \& {Vovk}}]{2019ICRC...36..739M}
{Mart{\'\i}nez-Huerta}, H., {Biteau}, J., {Lefaucheur}, J., {et~al.} 2019, in
  International Cosmic Ray Conference, Vol.~36, 36th International Cosmic Ray
  Conference (ICRC2019), 739

\bibitem[{{Mirzoyan}(2014)}]{2014ATel.6062....1M}
{Mirzoyan}, R. 2014, The Astronomer's Telegram, 6062, 1

\bibitem[{{Neronov} {et~al.}(2015){Neronov}, {Semikoz}, {Taylor}, \&
  {Vovk}}]{2015A&A...575A..21N}
{Neronov}, A., {Semikoz}, D., {Taylor}, A.~M., \& {Vovk}, I. 2015, \aap, 575,
  A21

\bibitem[{{Neronov} {et~al.}(2011){Neronov}, {Semikoz}, \&
  {Vovk}}]{2011A&A...529A..59N}
{Neronov}, A., {Semikoz}, D., \& {Vovk}, I. 2011, \aap, 529, A59

\bibitem[{{Neronov} \& {Semikoz}(2009)}]{2009PhRvD..80l3012N}
{Neronov}, A. \& {Semikoz}, D.~V. 2009, \prd, 80, 123012

\bibitem[{Neronov \& Semikoz(2012)}]{Neronov:2011kg}
Neronov, A. \& Semikoz, D.~V. 2012, Astrophys. J., 757, 61

\bibitem[{Neronov {et~al.}(2017)Neronov, Semikoz, \&
  Ptitsyna}]{Neronov:2016ksj}
Neronov, A., Semikoz, D.~V., \& Ptitsyna, K. 2017, Astron. Astrophys., 603,
  A135

\bibitem[{{Neronov} \& {Vovk}(2010)}]{2010Sci...328...73N}
{Neronov}, A. \& {Vovk}, I. 2010, Science, 328, 73

\bibitem[{Neronov \& Semikoz(2002)}]{Neronov:2002xv}
Neronov, A.~Y. \& Semikoz, D.~V. 2002, Phys. Rev. D, 66, 123003

\bibitem[{{Nilsson} {et~al.}(2024){Nilsson}, {Fallah Ramazani}, {Lindfors},
  {Goldoni}, {Becerra Gonz{\'a}lez}, {Acosta Pulido}, {Clavero},
  {Otero-Santos}, {Pursimo}, {Pita}, {Kouch}, {Boisson}, {Backes}, {Cotter},
  {D'Ammando}, \& {Kasai}}]{2024A&A...691A.154N}
{Nilsson}, K., {Fallah Ramazani}, V., {Lindfors}, E., {et~al.} 2024, \aap, 691,
  A154

\bibitem[{Oikonomou {et~al.}(2019)Oikonomou, Murase, Padovani, Resconi, \&
  M\'esz\'aros}]{Oikonomou:2019djc}
Oikonomou, F., Murase, K., Padovani, P., Resconi, E., \& M\'esz\'aros, P. 2019,
  Mon. Not. Roy. Astron. Soc., 489, 4347

\bibitem[{{Olmo-Garc{\'\i}a} {et~al.}(2022){Olmo-Garc{\'\i}a}, {Paliya},
  {{\'A}lvarez Crespo}, {Kumar}, {Dom{\'\i}nguez}, {Gil de Paz}, \&
  {S{\'a}nchez-Bl{\'a}zquez}}]{2022MNRAS.516.5702O}
{Olmo-Garc{\'\i}a}, A., {Paliya}, V.~S., {{\'A}lvarez Crespo}, N., {et~al.}
  2022, \mnras, 516, 5702

\bibitem[{Padovani {et~al.}(2022)Padovani, Boccardi, Falomo, \&
  Giommi}]{Padovani:2022wjk}
Padovani, P., Boccardi, B., Falomo, R., \& Giommi, P. 2022, Mon. Not. Roy.
  Astron. Soc., 511, 4697

\bibitem[{Padovani {et~al.}(2019)Padovani, Oikonomou, Petropoulou, Giommi, \&
  Resconi}]{Padovani:2019xcv}
Padovani, P., Oikonomou, F., Petropoulou, M., Giommi, P., \& Resconi, E. 2019,
  Mon. Not. Roy. Astron. Soc., 484, L104

\bibitem[{{Paiano} {et~al.}(2020){Paiano}, {Falomo}, {Treves}, \&
  {Scarpa}}]{2020MNRAS.497...94P}
{Paiano}, S., {Falomo}, R., {Treves}, A., \& {Scarpa}, R. 2020, \mnras, 497, 94

\bibitem[{{Paiano} {et~al.}(2017){Paiano}, {Landoni}, {Falomo}, {Treves},
  {Scarpa}, \& {Righi}}]{2017ApJ...837..144P}
{Paiano}, S., {Landoni}, M., {Falomo}, R., {et~al.} 2017, \apj, 837, 144

\bibitem[{Paiano {et~al.}(2017)Paiano, Landoni, Falomo, Treves, Scarpa, \&
  Righi}]{Paiano:2017pol}
Paiano, S., Landoni, M., Falomo, R., {et~al.} 2017, Astrophys. J., 837, 144

\bibitem[{Paliya {et~al.}(2021)Paliya, Dom\'\i{}nguez, Ajello,
  Olmo-Garc\'\i{}a, \& Hartmann}]{Paliya:2021tfi}
Paliya, V.~S., Dom\'\i{}nguez, A., Ajello, M., Olmo-Garc\'\i{}a, A., \&
  Hartmann, D. 2021, Astrophys. J. Suppl., 253, 46

\bibitem[{{Planck Collaboration} {et~al.}(2020){Planck Collaboration},
  {Aghanim}, {Akrami}, {Ashdown}, {Aumont}, {Baccigalupi}, {Ballardini},
  {Banday}, {Barreiro}, {Bartolo}, {Basak}, {Battye}, {Benabed}, {Bernard},
  {Bersanelli}, {Bielewicz}, {Bock}, {Bond}, {Borrill}, {Bouchet}, {Boulanger},
  {Bucher}, {Burigana}, {Butler}, {Calabrese}, {Cardoso}, {Carron},
  {Challinor}, {Chiang}, {Chluba}, {Colombo}, {Combet}, {Contreras}, {Crill},
  {Cuttaia}, {de Bernardis}, {de Zotti}, {Delabrouille}, {Delouis}, {Di
  Valentino}, {Diego}, {Dor{\'e}}, {Douspis}, {Ducout}, {Dupac}, {Dusini},
  {Efstathiou}, {Elsner}, {En{\ss}lin}, {Eriksen}, {Fantaye}, {Farhang},
  {Fergusson}, {Fernandez-Cobos}, {Finelli}, {Forastieri}, {Frailis},
  {Fraisse}, {Franceschi}, {Frolov}, {Galeotta}, {Galli}, {Ganga},
  {G{\'e}nova-Santos}, {Gerbino}, {Ghosh}, {Gonz{\'a}lez-Nuevo}, {G{\'o}rski},
  {Gratton}, {Gruppuso}, {Gudmundsson}, {Hamann}, {Handley}, {Hansen},
  {Herranz}, {Hildebrandt}, {Hivon}, {Huang}, {Jaffe}, {Jones}, {Karakci},
  {Keih{\"a}nen}, {Keskitalo}, {Kiiveri}, {Kim}, {Kisner}, {Knox},
  {Krachmalnicoff}, {Kunz}, {Kurki-Suonio}, {Lagache}, {Lamarre}, {Lasenby},
  {Lattanzi}, {Lawrence}, {Le Jeune}, {Lemos}, {Lesgourgues}, {Levrier},
  {Lewis}, {Liguori}, {Lilje}, {Lilley}, {Lindholm}, {L{\'o}pez-Caniego},
  {Lubin}, {Ma}, {Mac{\'\i}as-P{\'e}rez}, {Maggio}, {Maino}, {Mandolesi},
  {Mangilli}, {Marcos-Caballero}, {Maris}, {Martin}, {Martinelli},
  {Mart{\'\i}nez-Gonz{\'a}lez}, {Matarrese}, {Mauri}, {McEwen}, {Meinhold},
  {Melchiorri}, {Mennella}, {Migliaccio}, {Millea}, {Mitra},
  {Miville-Desch{\^e}nes}, {Molinari}, {Montier}, {Morgante}, {Moss}, {Natoli},
  {N{\o}rgaard-Nielsen}, {Pagano}, {Paoletti}, {Partridge}, {Patanchon},
  {Peiris}, {Perrotta}, {Pettorino}, {Piacentini}, {Polastri}, {Polenta},
  {Puget}, {Rachen}, {Reinecke}, {Remazeilles}, {Renzi}, {Rocha}, {Rosset},
  {Roudier}, {Rubi{\~n}o-Mart{\'\i}n}, {Ruiz-Granados}, {Salvati}, {Sandri},
  {Savelainen}, {Scott}, {Shellard}, {Sirignano}, {Sirri}, {Spencer},
  {Sunyaev}, {Suur-Uski}, {Tauber}, {Tavagnacco}, {Tenti}, {Toffolatti},
  {Tomasi}, {Trombetti}, {Valenziano}, {Valiviita}, {Van Tent}, {Vibert},
  {Vielva}, {Villa}, {Vittorio}, {Wandelt}, {Wehus}, {White}, {White},
  {Zacchei}, \& {Zonca}}]{2020A&A...641A...6P}
{Planck Collaboration}, {Aghanim}, N., {Akrami}, Y., {et~al.} 2020, \aap, 641,
  A6

\bibitem[{{Prandini} \& {Ghisellini}(2022)}]{2022Galax..10...35P}
{Prandini}, E. \& {Ghisellini}, G. 2022, Galaxies, 10, 35

\bibitem[{Sahakyan {et~al.}(2022)Sahakyan, Giommi, Padovani, Petropoulou,
  B\'egu\'e, Boccardi, \& Gasparyan}]{Sahakyan:2022nbz}
Sahakyan, N., Giommi, P., Padovani, P., {et~al.} 2022, Mon. Not. Roy. Astron.
  Soc., 519, 1396

\bibitem[{{Saldana-Lopez} {et~al.}(2021){Saldana-Lopez}, {Dom{\'\i}nguez},
  {P{\'e}rez-Gonz{\'a}lez}, {Finke}, {Ajello}, {Primack}, {Paliya}, \&
  {Desai}}]{2021MNRAS.507.5144S}
{Saldana-Lopez}, A., {Dom{\'\i}nguez}, A., {P{\'e}rez-Gonz{\'a}lez}, P.~G.,
  {et~al.} 2021, \mnras, 507, 5144

\bibitem[{Shaw {et~al.}(2013)Shaw, Romani, Cotter, Healey, Michelson, Readhead,
  Richards, Max-Moerbeck, King, \& Potter}]{Shaw:2013pp}
Shaw, M.~S., Romani, R.~W., Cotter, G., {et~al.} 2013, Astrophys. J., 764, 135

\bibitem[{Sol {et~al.}(2013)}]{CTAConsortium:2013tmf}
Sol, H. {et~al.} 2013, Astropart. Phys., 43, 215

\bibitem[{{Ulgiati} {et~al.}(2024){Ulgiati}, {Paiano}, {Treves}, {Falomo},
  {Sbarufatti}, {Pintore}, {Russell}, \& {Cusumano}}]{2024MNRAS.530.4626U}
{Ulgiati}, A., {Paiano}, S., {Treves}, A., {et~al.} 2024, \mnras, 530, 4626

\bibitem[{{V{\'e}ron-Cetty} \& {V{\'e}ron}(2010)}]{2010A&A...518A..10V}
{V{\'e}ron-Cetty}, M.~P. \& {V{\'e}ron}, P. 2010, \aap, 518, A10

\bibitem[{{Xiao} {et~al.}(2022){Xiao}, {Fan}, {Ouyang}, {Hu}, {Chen}, {Fu}, \&
  {Zhang}}]{2022ApJ...936..146X}
{Xiao}, H., {Fan}, J., {Ouyang}, Z., {et~al.} 2022, \apj, 936, 146

\bibitem[{{Zhang} {et~al.}(2025){Zhang}, {Yang}, {Zhang}, {Xie}, {Liu}, {Yin},
  {Wang}, {Ma}, \& {Cao}}]{2025ChPhC..49c5001Z}
{Zhang}, Z., {Yang}, R., {Zhang}, S., {et~al.} 2025, Chinese Physics C, 49,
  035001

\end{thebibliography}

\onecolumn

\begin{longtable}{lllllllllll}
\caption{IACT-detected high Galactic latitude VHE sources with time-average flux above $10^{-12}$~erg/(cm$^2$s). $N_{100}$ is the number of photons above 100~GeV within 68\% (95\%) containement radius of the PSF. $N_{300}$ is the number of photons above 300~GeV within the 95\% containement circle. Source types are BLLac -- BL Lacs type objects, RG -- radio galaxies. The "E" tad in the last column marks extreme blazars according to classification of Ref. \cite{2019A&A...632A..77C}. The VHE band flux $F_{100}$ is in the units of $10^{-12}$~erg/cm$^2$s. }\\
\label{tab:known_sources}
&4FGL               & RA & Dec & $N_{100}$  &$N_{300}$  & $\Gamma$  &$z$ & $F_{100}$ &Type &HSP  \\ 
\hline
\endfirsthead
\caption{continued.}\\
\hline
&4FGL               & RA & Dec & $N_{100}$  &$N_{300}$  & $\Gamma$   &$z$ & $F_{100}$ &Type & HSP  \\ 
\hline
\endhead
\hline
\endfoot
\hline
1  &  KUV 00311-1938  &  8.395  &  -19.359  &  3(7) &  1  &  1.751$\pm$0.018  &  0.61  & $1.9_{-0.6}^{+0.9}$  &  BLL  & \\
2  &  RGB J0136+391  &  24.142  &  39.1  &  6(7) &  0  &  1.72$\pm$0.013  &  0.75  & $1.7_{-0.5}^{+0.8}$  &  BLL  & \\
3  &  3C 66A  &  35.67  &  43.036  &  15(25) &  0  &  1.976$\pm$0.008  &  0.34  & $6.1_{-1.1}^{+1.4}$  &  BLL  & \\
4  &  1ES 0229+200  &  38.214  &  20.316  &  2(6) &  1  &  1.752$\pm$0.08  &  0.14  & $1.6_{-0.5}^{+0.8}$  &  BLL  & E \\
5  &  1RXS J023832.6-311658  &  39.621  &  -31.283  &  4(6) &  1  &  1.821$\pm$0.037  &  0.233  & $1.5_{-0.5}^{+0.8}$  &  BLL  & \\
6  &  PKS 0301-243  &  45.862  &  -24.122  &  8(10) &  0  &  1.912$\pm$0.013  &  0.26  & $2.6_{-0.7}^{+1.0}$  &  BLL  & \\
7  &  IC 310  &  49.215  &  41.346  &  3(5) &  1  &  1.882$\pm$0.124  &  0.019  & $1.2_{-0.4}^{+0.7}$  &  RDG  & \\
8  &  NGC 1275  &  49.958  &  41.512  &  11(12) &  1  &  2.123$\pm$0.005  &  0.017  & $3.0_{-0.7}^{+1.0}$  &  RDG  & \\
9  &  RBS 0413  &  49.972  &  18.753  &  3(7) &  1  &  1.709$\pm$0.054  &  0.19  & $1.8_{-0.6}^{+0.8}$  &  BLL  & E \\
10  &  PKS 0346-27  &  57.154  &  -27.822  &  3(3) &  0  &  2.086$\pm$0.006  &  0.99  & $0.8_{-0.3}^{+0.6}$  &  FSRQ  & \\
11  &  1ES 0347-121  &  57.354  &  -11.994  &  3(3) &  1  &  1.854$\pm$0.063  &  0.188  & $0.8_{-0.3}^{+0.6}$  &  BLL  & E \\
12  &  1ES 0414+009  &  64.227  &  1.088  &  3(7) &  2  &  1.849$\pm$0.054  &  0.287  & $1.9_{-0.6}^{+0.9}$  &  BLL  & \\
13  &  PKS 0447-439  &  72.358  &  -43.835  &  22(38) &  1  &  1.869$\pm$0.008  &  0.107  & $9.2_{-1.4}^{+1.6}$  &  BLL  & \\
14  &  1ES 0502+675  &  76.996  &  67.622  &  25(34) &  6  &  1.589$\pm$0.02  &  0.314  & $7.1_{-1.1}^{+1.3}$  &  BLL  & E \\
15  &  TXS 0506+056  &  77.359  &  5.701  &  4(6) &  0  &  2.108$\pm$0.012  &  0.336  & $1.6_{-0.5}^{+0.8}$  &  BLL  & \\
16  &  PKS 0548-322  &  87.625  &  -32.277  &  4(5) &  0  &  1.829$\pm$0.071  &  0.069  & $1.2_{-0.4}^{+0.7}$  &  BLL  & E \\
17  &  PKS 0625-35  &  96.775  &  -35.487  &  9(11) &  3  &  1.913$\pm$0.029  &  0.055  & $2.7_{-0.7}^{+0.9}$  &  RDG  & \\
18  &  1ES 0647+250  &  102.699  &  25.055  &  15(28) &  2  &  1.738$\pm$0.014  &  0.203  & $7.4_{-1.3}^{+1.5}$  &  BLL  & \\
19  &  RGB J0710+591  &  107.623  &  59.135  &  3(4) &  1  &  1.703$\pm$0.043  &  0.125  & $0.9_{-0.4}^{+0.6}$  &  BLL  & E \\
20  &  S5 0716+714  &  110.488  &  71.34  &  13(19) &  1  &  2.065$\pm$0.005  &  0.3  & $3.8_{-0.8}^{+1.0}$  &  BLL  & \\
21  &  PGC 2402248  &  113.362  &  51.88  &  3(5) &  1  &  1.769$\pm$0.079  &  0.065  & $1.2_{-0.4}^{+0.7}$  &  BLL\footnote{we changed BCU to BLL based on the source type in SIMBAD}  & E \\
22  &  1ES 0806+524  &  122.462  &  52.314  &  14(20) &  2  &  1.882$\pm$0.014  &  0.137  & $4.8_{-1.0}^{+1.2}$  &  BLL  & \\
23  &  1RXS J081201.8+023735  &  123.009  &  2.628  &  3(3) &  1  &  1.882$\pm$0.068  &  0.2  & $0.8_{-0.4}^{+0.6}$  &  BLL  & \\
24  &  MRC 0910-208  &  138.227  &  -21.045  &  6(11) &  1  &  1.816$\pm$0.038  &  0.198  & $2.9_{-0.7}^{+1.0}$  &  BLL  & E \\
25  &  S4 0954+65  &  149.69  &  65.568  &  4(5) &  0  &  2.191$\pm$0.01  &  0.367  & $1.0_{-0.4}^{+0.6}$  &  BLL  & \\
26  &  1RXS J101015.9-311909  &  152.572  &  -31.321  &  6(6) &  2  &  1.765$\pm$0.051  &  0.143  & $1.5_{-0.5}^{+0.8}$  &  BLL  & \\
27  &  1ES 1011+496  &  153.768  &  49.434  &  19(31) &  3  &  1.85$\pm$0.008  &  0.212  & $7.4_{-1.2}^{+1.4}$  &  BLL  & \\
28  &  1ES 1101-232  &  165.909  &  -23.496  &  1(6) &  1  &  1.699$\pm$0.06  &  0.186  & $1.6_{-0.5}^{+0.8}$  &  BLL  & E \\
29  &  Markarian 421  &  166.119  &  38.207  &  232(314) &  69  &  1.785$\pm$0.004  &  0.03  & $77.6_{-4.3}^{+4.5}$  &  BLL  & \\
30  &  RX J1136.5+6737  &  174.118  &  67.613  &  3(4) &  2  &  1.775$\pm$0.043  &  0.134  & $0.8_{-0.3}^{+0.5}$  &  BLL  & E \\
31  &  Markarian 180  &  174.122  &  70.154  &  6(8) &  3  &  1.828$\pm$0.025  &  0.046  & $1.6_{-0.5}^{+0.7}$  &  BLL  & \\
32  &  1ES 1215+303  &  184.476  &  30.118  &  10(16) &  2  &  1.936$\pm$0.01  &  0.131\footnote{From \citet{2019AJ....157...41F}.}   & $4.0_{-0.9}^{+1.1}$  &  BLL  & \\
33  &  1ES 1218+304  &  185.345  &  30.168  &  28(39) &  10  &  1.737$\pm$0.017  &  0.184  & $9.6_{-1.4}^{+1.7}$  &  BLL  & \\
34  &  W Comae  &  185.378  &  28.238  &  2(3) &  1  &  2.156$\pm$0.017  &  0.103  & $0.7_{-0.3}^{+0.6}$  &  BLL  & \\
35  &  MS 1221.8+2452  &  186.116  &  24.614  &  3(3) &  0  &  1.937$\pm$0.035  &  0.219  & $0.8_{-0.3}^{+0.6}$  &  BLL  & \\
36  &  S3 1227+25  &  187.56  &  25.298  &  2(3) &  0  &  2.096$\pm$0.013  &  0.135  & $0.8_{-0.3}^{+0.6}$  &  BLL  & \\
37  &  M 87  &  187.712  &  12.388  &  5(7) &  0  &  2.064$\pm$0.028  &  0.004  & $1.9_{-0.6}^{+0.8}$  &  RDG  & \\
38  &  3C 279  &  194.042  &  -5.789  &  2(3) &  0  &  2.293$\pm$0.004  &  0.535  & $0.8_{-0.4}^{+0.6}$  &  FSRQ  & \\
39  &  OP 313  &  197.632  &  32.355  &  3(3) &  0  &  2.23$\pm$0.014  &  0.996  & $0.7_{-0.3}^{+0.6}$  &  FSRQ  & \\
40  &  1ES 1312-423  &  198.768  &  -42.611  &  5(8) &  1  &  1.794$\pm$0.066  &  0.105  & $2.1_{-0.6}^{+0.9}$  &  BLL  & E \\
41  &  Centaurus A  &  201.381  &  -43.016  &  5(6) &  1  &  2.574$\pm$0.019  &  0.002  & $1.6_{-0.5}^{+0.8}$  &  RDG  & \\
42  &  PKS 1413+135  &  214.003  &  13.339  &  3(3) &  0  &  2.069$\pm$0.021  &  0.247  & $0.8_{-0.3}^{+0.6}$  &  BLL\footnote{BCU in Fermi LAT catalog, BLL in \cite{2010A&A...518A..10V} catalog.}  & \\
43  &  RBS 1366  &  214.494  &  25.724  &  4(5) &  1  &  1.473$\pm$0.069  &  0.236  & $1.2_{-0.4}^{+0.7}$  &  BLL  & E \\
44  &  PKS 1424+240  &  216.756  &  23.801  &  16(31) &  0  &  1.816$\pm$0.008  &  0.16  & $7.4_{-1.2}^{+1.5}$  &  BLL  & \\
45  &  H 1426+428  &  217.129  &  42.678  &  9(11) &  4  &  1.659$\pm$0.038  &  0.129  & $2.5_{-0.6}^{+0.9}$  &  BLL  & E \\
46  &  PKS 1440-389  &  220.991  &  -39.148  &  18(21) &  1  &  1.834$\pm$0.015  &  0.065  & $5.7_{-1.1}^{+1.4}$  &  BLL  & \\
47  &  PKS 1510-089  &  228.215  &  -9.106  &  6(7) &  0  &  2.379$\pm$0.005  &  0.356  & $1.9_{-0.6}^{+0.9}$  &  FSRQ  & \\
48  &  AP Lib  &  229.425  &  -24.373  &  4(9) &  2  &  2.122$\pm$0.011  &  0.049  & $2.5_{-0.7}^{+1.0}$  &  BLL  & \\
49  &  TXS 1515-273  &  229.512  &  -27.531  &  2(3) &  0  &  2.036$\pm$0.033  &  0.14  & $0.8_{-0.4}^{+0.6}$  &  BLL  & \\
50  &  PG 1553+113  &  238.931  &  11.188  &  65(88) &  3  &  1.697$\pm$0.007  &  0.36  & $21.9_{-2.2}^{+2.5}$  &  BLL  & \\
51  &  Markarian 501  &  253.474  &  39.759  &  81(109) &  22  &  1.795$\pm$0.008  &  0.034  & $23.5_{-2.1}^{+2.4}$  &  BLL  & E \\
52  &  H 1722+119  &  261.271  &  11.875  &  8(10) &  2  &  1.869$\pm$0.019  &  0.018  & $2.5_{-0.7}^{+0.9}$  &  BLL  & \\
53  &  1ES 1727+502  &  262.078  &  50.227  &  5(11) &  4  &  1.785$\pm$0.022  &  0.055  & $2.3_{-0.6}^{+0.8}$  &  BLL  & E \\
54  &  1ES 1741+196  &  266.008  &  19.596  &  3(4) &  0  &  1.922$\pm$0.043  &  0.08  & $0.9_{-0.4}^{+0.6}$  &  BLL  & E \\
55  &  B2 1811+31  &  273.387  &  31.75  &  2(3) &  1  &  2.009$\pm$0.026  &  0.117  & $0.7_{-0.3}^{+0.5}$  &  BLL  & \\
56  &  1RXS J195815.6-301119  &  299.581  &  -30.181  &  1(6) &  0  &  1.833$\pm$0.041  &  0.119  & $1.7_{-0.6}^{+0.8}$  &  BLL  & E \\
57  &  1ES 1959+650  &  300.011  &  65.148  &  89(115) &  32  &  1.815$\pm$0.009  &  0.047  & $22.7_{-2.0}^{+2.2}$  &  BLL  & \\
58  &  PKS 2005-489  &  302.359  &  -48.825  &  16(23) &  1  &  1.841$\pm$0.017  &  0.071  & $6.3_{-1.2}^{+1.5}$  &  BLL  & \\
59  &  MG2 J204208+2426  &  310.536  &  24.458  &  4(5) &  1  &  1.947$\pm$0.052  &  0.104  & $1.2_{-0.4}^{+0.7}$  &  BLL  & E \\
60  &  PKS 2155-304  &  329.714  &  -30.225  &  58(83) &  8  &  1.857$\pm$0.006  &  0.116  & $23.3_{-2.4}^{+2.7}$  &  BLL  & \\
61  &  BL Lacertae  &  330.695  &  42.282  &  30(40) &  5  &  2.134$\pm$0.004  &  0.066  & $9.1_{-1.3}^{+1.6}$  &  BLL  & \\
62  &  RGB J2243+203  &  340.99  &  20.357  &  3(4) &  0  &  1.878$\pm$0.015  &  0.39  & $1.0_{-0.4}^{+0.6}$  &  BLL  & \\
63  &  B3 2247+381  &  342.514  &  38.425  &  5(7) &  2  &  1.77$\pm$0.048  &  0.119  & $1.6_{-0.5}^{+0.7}$  &  BLL  & \\
64  &  1ES 2322-409  &  351.182  &  -40.683  &  3(4) &  0  &  1.762$\pm$0.036  &  0.174  & $1.1_{-0.4}^{+0.7}$  &  BLL  & \\
65  &  H 2356-309  &  359.772  &  -30.637  &  3(5) &  0  &  1.838$\pm$0.059  &  0.165  & $1.4_{-0.5}^{+0.8}$  &  BLL  & E \\

\hline
\end{longtable}

\begin{longtable}{lllllllllll}
\caption{New VHE detections at $4\sigma$. Bold face marks the sources detected at $>5\sigma$ in the VHE band alone.  Notations are the same as in Table \ref{tab:known_sources}.}\\
\label{tab:new_sources}
&4FGL               & RA & Dec & $N_{100}$  &$N_{300}$  & $\Gamma$   &$z$ & $F_{100}$ &Type & HSP \\ 
\hline
\endfirsthead
\caption{continued.}\\
\hline
&4FGL               & RA & Dec & $N_{100}$  &$N_{300}$  & $\Gamma$   &$z$ & $F_{100}$ &Type & HSP  \\ 
\hline
\endhead
\hline
\endfoot
\hline
1  &  {\bf RX J0022.0+0006} &  5.515  &  0.113  &  5(9) &  1  &  1.66$\pm$0.12  &  0.306  & $2.5_{-0.7}^{+1.0}$  &  BLL  &  \\
2  &  PKS 0048-09  &  12.675  &  -9.494  &  2(4) &  0  &  2.01$\pm$0.02  &  0.635  & $1.1_{-0.4}^{+0.7}$  &  BLL  &  \\
3  &  1RXS J005117.7-624154  &  12.824  &  -62.704  &  2(5) &  1  &  1.77$\pm$0.02  &  0.156  & $1.3_{-0.5}^{+0.7}$  &  BLL  &  \\
4  & {\bf  RX J0115.7+2519}  &  18.954  &  25.332  &  3(5) &  0  &  1.91$\pm$0.03  &  0.375  & $1.3_{-0.5}^{+0.7}$  &  BLL  &  \\
5  &  PKS 0118-272  &  20.123  &  -27.023  &  2(5) &  0  &  1.91$\pm$0.02  &  0.557  & $1.3_{-0.5}^{+0.7}$  &  BLL  &  \\
6  &  {\bf  1RXS J012338.2-231100}  &  20.938  &  -23.194  &  4(4) &  0  &  1.85$\pm$0.05  &  0.404  & $1.1_{-0.4}^{+0.7}$  &  BLL  & E \\
7  &  PKS 0139-09  &  25.363  &  -9.483  &  3(3) &  0  &  2.12$\pm$0.02  &  0.5  & $0.8_{-0.4}^{+0.6}$  &  BLL  &  \\
8  & {\bf  1RXS J015308.4+751756}  &  28.253  &  75.288  &  4(7) &  0  &  2.0$\pm$0.07  &  -1  & $1.4_{-0.4}^{+0.6}$  &  BLL  &  \\
9  & {\bf  PMN J0209-5229 } &  32.349  &  -52.48  &  7(12) &  0  &  1.84$\pm$0.02  &  0.12  & $3.0_{-0.7}^{+1.0}$  &  BLL  &  \\
10  & {\bf  4FGL J0211.2+1051}  &  32.809  &  10.857  &  3(5) &  1  &  2.08$\pm$0.01  &  0.2  & $1.3_{-0.5}^{+0.7}$  &  BLL  &  \\
11  & {\bf  RX J0212.3-0222 } &  33.066  &  -2.319  &  4(5) &  1  &  1.97$\pm$0.08  &  0.25  & $1.4_{-0.5}^{+0.8}$  &  BLL  & E \\
12  &  RBS 0298  &  34.141  &  23.229  &  3(3) &  0  &  1.88$\pm$0.09  &  0.288  & $0.8_{-0.3}^{+0.6}$  &  BLL  &  \\
13  &  TXS 0230+067  &  38.386  &  6.908  &  2(3) &  0  &  2.32$\pm$0.06  &  -1  & $0.8_{-0.4}^{+0.6}$  &  BLL\footnote{We changed from BCU to BLL following \cite{2022MNRAS.516.5702O}.}   &  \\
14  & {\bf  RBS 0351}  &  41.158  &  -58.326  &  5(6) &  0  &  1.72$\pm$0.04  &  0.265  & $1.5_{-0.5}^{+0.7}$  &  BLL  & E \\
15  & {\bf  V Zw 326}  &  48.228  &  36.234  &  6(6) &  3  &  1.87$\pm$0.07  &  0.071  & $1.5_{-0.5}^{+0.8}$  &  BLL  &  \\
16  & {\bf  GB6 J0316+0904 } &  49.058  &  9.095  &  3(4) &  1  &  1.89$\pm$0.03  &  0.372  & $1.1_{-0.4}^{+0.7}$  &  BLL  &  \\
17  &  1RXS J032521.8-563543  &  51.379  &  -56.591  &  3(3) &  0  &  1.91$\pm$0.03  &  0.06  & $0.7_{-0.3}^{+0.6}$  &  BLL  &  \\
18  &  RBS 0421  &  51.418  &  -16.781  &  2(3) &  0  &  1.89$\pm$0.04  &  0.291  & $0.8_{-0.3}^{+0.6}$  &  BLL  &  \\
19  &  1H 0323+022  &  51.572  &  2.423  &  2(3) &  0  &  1.92$\pm$0.05  &  0.147  & $0.8_{-0.4}^{+0.6}$  &  BLL  &  \\
20  &  4C +39.12  &  53.582  &  39.339  &  2(4) &  1  &  1.82$\pm$0.1  &  0.02  & $1.0_{-0.4}^{+0.6}$  &  RDG  &  \\
21  &  1RXS J033623.3-034727  &  54.13  &  -3.812  &  3(3) &  0  &  1.68$\pm$0.12  &  0.162  & $0.8_{-0.4}^{+0.6}$  &  BLL  & E \\
22  &  NVSS J033859-284619  &  54.745  &  -28.801  &  2(3) &  0  &  1.97$\pm$0.05  &  0.27  & $0.8_{-0.3}^{+0.6}$  &  BCU  &  \\
23  &  PKS 0336-177  &  54.812  &  -17.6  &  2(3) &  1  &  1.92$\pm$0.04  &  0.066  & $0.8_{-0.3}^{+0.6}$  &  BLL  &  \\
24  &  4FGL J0350.0+0640  &  57.504  &  6.675  &  3(3) &  0  &  1.47$\pm$0.18  &  0.26  & $0.8_{-0.4}^{+0.6}$  &  BCU  & E \\
25  &  PKS 0352-686  &  58.274  &  -68.528  &  2(5) &  0  &  1.67$\pm$0.07  &  0.088  & $1.3_{-0.5}^{+0.7}$  &  BLL  & E \\
26  &  PKS 0426-380  &  67.173  &  -37.94  &  3(3) &  0  &  2.1$\pm$0.01  &  1.111  & $0.7_{-0.3}^{+0.6}$  &  BLL  &  \\
27  &  RBS 0589  &  72.152  &  -16.548  &  3(3) &  0  &  1.88$\pm$0.04  &  -1  & $0.8_{-0.3}^{+0.6}$  &  BLL  & E \\
28  &  RX J0505.9+6113  &  76.512  &  61.226  &  2(3) &  0  &  1.96$\pm$0.05  &  0.27  & $0.7_{-0.3}^{+0.5}$  &  BLL  &  \\
29  &  1ES 0505-546  &  76.748  &  -54.588  &  2(3) &  0  &  1.65$\pm$0.04  &  -1  & $0.7_{-0.3}^{+0.6}$  &  BLL  &  \\
30  &  RBS 0625  &  77.488  &  -64.289  &  3(3) &  0  &  1.97$\pm$0.06  &  -1  & $0.7_{-0.3}^{+0.6}$  &  BLL  & E \\
31  &  GB6 J0516+7350  &  79.104  &  73.85  &  2(3) &  0  &  1.88$\pm$0.05  &  0.251  & $0.6_{-0.3}^{+0.5}$  &  BLL  &  \\
32  &  PKS 0537-441  &  84.709  &  -44.086  &  2(4) &  0  &  2.14$\pm$0.01  &  0.894  & $1.0_{-0.4}^{+0.6}$  &  BLL  &  \\
33  &  GB6 J0540+5823  &  85.148  &  58.388  &  3(3) &  0  &  1.88$\pm$0.05  &  -1  & $0.7_{-0.3}^{+0.5}$  &  BLL  &  \\
34  &  {\bf 1RXS J054357.3-553206 } &  85.981  &  -55.533  &  17(22) &  1  &  1.78$\pm$0.02  &  0.273  & $5.3_{-1.0}^{+1.2}$  &  BLL  &  \\
35  &  {\bf PMN J0622-2605}  &  95.596  &  -26.096  &  6(7) &  0  &  1.89$\pm$0.02  &  -1  & $1.7_{-0.5}^{+0.8}$  &  BLL  &  \\
36  &  {\bf GB6 J0723+2051 } &  110.949  &  20.837  &  4(6) &  1  &  1.82$\pm$0.08  &  0.21  & $1.6_{-0.5}^{+0.8}$  &  BLL  &  \\
37  &  PKS 0735+17  &  114.539  &  17.707  &  2(4) &  0  &  2.03$\pm$0.01  &  0.45  & $1.1_{-0.4}^{+0.7}$  &  BLL  &  \\
38  &  {\bf MS 0737.9+7441}  &  116.035  &  74.578  &  3(5) &  0  &  1.81$\pm$0.06  &  0.314  & $1.0_{-0.4}^{+0.6}$  &  BLL  &  \\
39  & {\bf  PMN J0746-4755}  &  116.669  &  -47.916  &  5(5) &  0  &  1.95$\pm$0.03  &  -1  & $1.2_{-0.4}^{+0.7}$  &  BLL  &  \\
40  & {\bf  RX J0805.4+7534 } &  121.362  &  75.577  &  13(15) &  4  &  1.84$\pm$0.02  &  0.121  & $2.9_{-0.7}^{+0.9}$  &  BLL  &  \\
41  &  {\bf  PMN J0816-1311 } &  124.112  &  -13.197  &  4(4) &  0  &  1.84$\pm$0.03  &  0.046  & $1.1_{-0.4}^{+0.7}$  &  BLL  &  \\
42  & {\bf  PMN J0847-2337 } &  131.757  &  -23.614  &  7(11) &  1  &  1.92$\pm$0.03  &  0.059  & $2.8_{-0.7}^{+1.0}$  &  BLL  &  \\
43  &  Ton 1015  &  137.651  &  33.491  &  2(3) &  0  &  1.94$\pm$0.03  &  0.354  & $0.8_{-0.3}^{+0.6}$  &  BLL  &  \\
44  &  RX J0912.5+1555  &  138.137  &  15.934  &  3(3) &  1  &  1.79$\pm$0.09  &  0.085  & $0.8_{-0.3}^{+0.6}$  &  BLL  & E \\
45  &  Ton 0396  &  138.986  &  29.553  &  2(5) &  0  &  1.86$\pm$0.02  &  1.521  & $1.3_{-0.5}^{+0.7}$  &  BLL  &  \\
46  & {\bf  4FGL J0934.5-1720}  &  143.628  &  -17.339  &  4(4) &  0  &  1.9$\pm$0.14  &  0.25  & $1.1_{-0.4}^{+0.7}$  &  BLL  & E \\
47  & {\bf  RBS 0804 } &  148.063  &  75.055  &  4(4) &  1  &  1.53$\pm$0.11  &  0.179  & $0.8_{-0.3}^{+0.5}$  &  BLL  & E \\
48  &  4C +55.17  &  149.416  &  55.384  &  3(3) &  0  &  2.02$\pm$0.01  &  0.903  & $0.7_{-0.3}^{+0.5}$  &  FSRQ  &  \\
49  &  NVSS J102703+060934  &  156.732  &  6.143  &  2(3) &  0  &  1.74$\pm$0.11  &  0.449  & $0.8_{-0.4}^{+0.6}$  &  BLL  &  \\
50  & {\bf  S5 1027+74}  &  157.792  &  74.702  &  3(5) &  1  &  2.18$\pm$0.07  &  0.123  & $1.0_{-0.3}^{+0.5}$  &  BLL  &  \\
51  & {\bf  1ES 1028+511}  &  157.845  &  50.884  &  6(9) &  0  &  1.73$\pm$0.03  &  0.36  & $2.1_{-0.6}^{+0.8}$  &  BLL  &  \\
52  & {\bf  GB6 J1037+5711}  &  159.429  &  57.192  &  8(9) &  0  &  1.89$\pm$0.01  &  0.831  & $2.0_{-0.6}^{+0.8}$  &  BLL  &  \\
53  &  RX J1055.5-0126  &  163.889  &  -1.432  &  2(3) &  0  &  1.87$\pm$0.07  &  0.33  & $0.8_{-0.4}^{+0.6}$  &  BLL  &  \\
54  & {\bf  RBS 0958}  &  169.271  &  20.229  &  6(6) &  0  &  1.9$\pm$0.03  &  0.138  & $1.6_{-0.5}^{+0.8}$  &  BLL  &  \\
55  & {\bf  RBS 0970}  &  170.201  &  42.204  &  10(12) &  0  &  1.64$\pm$0.02  &  0.124  & $2.9_{-0.7}^{+1.0}$  &  BLL  &  \\
56  &  PMN J1203-3926  &  180.852  &  -39.426  &  2(5) &  0  &  1.84$\pm$0.06  &  0.28  & $1.3_{-0.5}^{+0.7}$  &  BLL  &  \\
57  & {\bf  Ton 116}  &  190.812  &  36.458  &  4(6) &  0  &  1.78$\pm$0.02  &  1.066  & $1.4_{-0.5}^{+0.7}$  &  BLL  &  \\
58  &  PG 1246+586  &  192.084  &  58.343  &  2(3) &  0  &  1.92$\pm$0.01  &  0.847  & $0.6_{-0.3}^{+0.5}$  &  BLL  &  \\
59  &  4FGL J1310.6+2449  &  197.652  &  24.832  &  3(3) &  0  &  2.0$\pm$0.09  &  -1  & $0.7_{-0.3}^{+0.6}$  &  BLL  &  \\
60  & {\bf  SDSS J134105.10+395945.4}  &  205.321  &  39.974  &  4(6) &  2  &  1.79$\pm$0.08  &  0.171  & $1.4_{-0.5}^{+0.7}$  &  BLL  & E \\
61  &  PMN J1359-3746  &  209.967  &  -37.768  &  2(3) &  2  &  2.06$\pm$0.05  &  0.334  & $0.8_{-0.4}^{+0.6}$  &  BLL  &  \\
62  & {\bf  NVSS J141826-023336}  &  214.606  &  -2.559  &  4(4) &  0  &  1.79$\pm$0.02  &  0.0  & $1.1_{-0.4}^{+0.7}$  &  BLL  &  \\
63  & {\bf  1RXS J144037.4-384658}  &  220.155  &  -38.78  &  5(5) &  0  &  1.74$\pm$0.05  &  0.27  & $1.4_{-0.5}^{+0.8}$  &  BLL  & E \\
64  & {\bf  RBS 1432 } &  222.017  &  36.134  &  4(5) &  0  &  1.83$\pm$0.02  &  1.508  & $1.1_{-0.4}^{+0.6}$  &  BLL  &  \\
65  &  PMN J1506+0814  &  226.674  &  8.226  &  2(3) &  0  &  1.8$\pm$0.04  &  0.376  & $0.8_{-0.3}^{+0.6}$  &  BLL  &  \\
66  & {\bf  1ES 1544+820}  &  235.047  &  81.922  &  4(5) &  0  &  1.71$\pm$0.03  &  -1  & $0.9_{-0.3}^{+0.5}$  &  BLL  &  \\
67  & {\bf  NVSS J154203-291509}  &  235.5  &  -29.252  &  3(4) &  0  &  1.84$\pm$0.06  &  -1  & $1.1_{-0.4}^{+0.7}$  &  BLL  &  \\
68  & {\bf  NVSS J154419-064913}  &  236.079  &  -6.826  &  6(7) &  1  &  1.87$\pm$0.04  &  0.172  & $1.9_{-0.6}^{+0.9}$  &  BCU  &  \\
69  & {\bf  PMN J1548-2251} &  237.201  &  -22.847  &  5(8) &  1  &  1.8$\pm$0.04  &  0.192  & $2.2_{-0.7}^{+0.9}$  &  BLL  &  \\
70  & {\bf  1RXS J155333.4-311841}  &  238.392  &  -31.311  &  3(4) &  0  &  1.95$\pm$0.03  &  0.132  & $1.1_{-0.4}^{+0.7}$  &  BLL  &  \\
71  &  4FGL J1554.2+2008  &  238.553  &  20.148  &  2(3) &  0  &  1.82$\pm$0.09  &  0.222  & $0.7_{-0.3}^{+0.5}$  &  BLL  & E \\
72  & {\bf  PMN J1610-6649}  &  242.692  &  -66.815  &  4(5) &  0  &  1.76$\pm$0.02  &  0.11  & $1.4_{-0.5}^{+0.8}$  &  BLL  &  \\
73  &  NVSS J164328-064619  &  250.883  &  -6.775  &  2(3) &  0  &  2.17$\pm$0.13  &  0.082  & $0.8_{-0.4}^{+0.6}$  &  BLL  &  \\
74  &  1RXS J164602.3-094113  &  251.512  &  -9.713  &  2(3) &  0  &  2.12$\pm$0.14  &  -1  & $0.8_{-0.4}^{+0.6}$  &  BCU  &  \\
75  & {\bf  1RXS J171405.2-202747}  &  258.522  &  -20.486  &  2(6) &  1  &  1.61$\pm$0.1  &  0.09  & $1.6_{-0.5}^{+0.8}$  &  BCU  &  \\
76  &  {\bf TXS 1742-078}  &  266.364  &  -7.889  &  3(4) &  0  &  2.06$\pm$0.04  &  -1  & $1.1_{-0.4}^{+0.7}$  &  BLL  &  \\
77  &  {\bf S4 1749+70}  &  267.158  &  70.097  &  3(4) &  0  &  2.02$\pm$0.01  &  0.77  & $0.8_{-0.3}^{+0.5}$  &  BLL  &  \\
78  &  {\bf RX J1756.1+5522}  &  269.08  &  55.38  &  3(4) &  0  &  1.84$\pm$0.06  &  -1  & $0.8_{-0.3}^{+0.5}$  &  BLL  &  \\
79  &  S5 1803+784  &  270.173  &  78.467  &  2(3) &  0  &  2.24$\pm$0.01  &  0.68  & $0.6_{-0.2}^{+0.4}$  &  BLL  &  \\
80  &  NVSS J181118+034113  &  272.826  &  3.679  &  2(4) &  1  &  1.86$\pm$0.04  &  -1  & $1.0_{-0.4}^{+0.7}$  &  BLL  &  \\
81  &  NVSS J183632+195047  &  279.126  &  19.811  &  2(3) &  0  &  2.09$\pm$0.08  &  -1  & $0.7_{-0.3}^{+0.5}$  &  BLL\footnote{We changed from BCU to BLL following \cite{2022MNRAS.516.5702O}.}  &  \\
82  &  {\bf GB6 J1838+4802}  &  279.714  &  48.041  &  3(8) &  1  &  1.85$\pm$0.02  &  0.3  & $1.7_{-0.5}^{+0.7}$  &  BLL  &  \\
83  &  1RXS J184230.6-584202  &  280.611  &  -58.68  &  2(3) &  0  &  1.73$\pm$0.09  &  0.33  & $0.8_{-0.4}^{+0.6}$  &  BLL  & E \\
84  &  {\bf NVSS J185023+263151}  &  282.64  &  26.526  &  3(5) &  0  &  1.69$\pm$0.1  &  0.22  & $1.1_{-0.4}^{+0.6}$  &  BLL  &  \\
85  &  {\bf 4FGL J1904.1+3627}  &  286.034  &  36.453  &  3(4) &  0  &  1.85$\pm$0.06  &  0.078  & $0.9_{-0.3}^{+0.6}$  &  BLL  & E \\
86  &  PMN J1911-1908  &  287.868  &  -19.149  &  2(4) &  0  &  1.87$\pm$0.07  &  0.16  & $1.1_{-0.4}^{+0.7}$  &  BLL  & E \\
87  &  {\bf 1H 1914-194}  &  289.438  &  -19.363  &  10(12) &  2  &  1.95$\pm$0.02  &  0.137  & $3.3_{-0.8}^{+1.1}$  &  BLL  &  \\
88  &  {\bf PMN J1921-1607}  &  290.463  &  -16.123  &  3(4) &  0  &  1.83$\pm$0.04  &  -1  & $1.1_{-0.4}^{+0.7}$  &  BLL  &  \\
89  &  87GB 192614.4+614823  &  291.71  &  61.915  &  3(3) &  0  &  1.9$\pm$0.03  &  -1  & $0.6_{-0.3}^{+0.5}$  &  BLL  &  \\
90  &  PMN J1936-4719  &  294.242  &  -47.34  &  3(3) &  0  &  1.82$\pm$0.05  &  0.265  & $0.8_{-0.4}^{+0.6}$  &  BLL  & E \\
91  & {\bf  1RXS J194455.3-214318}  &  296.229  &  -21.722  &  3(6) &  0  &  1.84$\pm$0.04  &  0.28  & $1.7_{-0.6}^{+0.8}$  &  BLL  & E \\
92  &  S5 2007+77  &  301.393  &  77.883  &  3(3) &  1  &  2.23$\pm$0.02  &  0.342  & $0.6_{-0.2}^{+0.4}$  &  BLL  &  \\
93  &  PMN J2016-0903  &  304.098  &  -9.062  &  2(4) &  1  &  1.98$\pm$0.04  &  0.367  & $1.1_{-0.4}^{+0.7}$  &  BLL  &  \\
94  &  4FGL J2026.1+7645  &  306.538  &  76.759  &  2(4) &  0  &  1.73$\pm$0.08  &  0.29  & $0.8_{-0.3}^{+0.5}$  &  BLL\footnote{We changed from BCU to BLL following \cite{2017A&A...602A..86L}}  &  \\
95  &  {\bf NVSS J204150-373341}  &  310.479  &  -37.587  &  3(5) &  1  &  1.88$\pm$0.08  &  0.099  & $1.4_{-0.5}^{+0.8}$  &  BLL  & E \\
96  &  1RXS J204921.6-003930  &  312.446  &  -0.616  &  2(3) &  1  &  1.53$\pm$0.25  &  0.257  & $0.8_{-0.4}^{+0.6}$  &  BLL  &  \\
97  &  1RXS J205528.2-002123  &  313.853  &  -0.34  &  3(3) &  1  &  1.8$\pm$0.07  &  0.407  & $0.8_{-0.4}^{+0.6}$  &  BLL  & E \\
98  &  NVSS J210421-021239  &  316.079  &  -2.209  &  3(3) &  0  &  1.66$\pm$0.09  &  -1  & $0.8_{-0.4}^{+0.6}$  &  BLL  &  \\
99  &  4FGL J2139.4-4235  &  324.855  &  -42.59  &  2(5) &  0  &  2.01$\pm$0.01  &  -1  & $1.4_{-0.5}^{+0.8}$  &  BLL  &  \\
100  &  NVSS J214637-134359  &  326.645  &  -13.735  &  2(3) &  0  &  1.76$\pm$0.03  &  -1  & $0.8_{-0.4}^{+0.7}$  &  BLL  &  \\
101  &  RBS 1792  &  328.282  &  -0.693  &  3(3) &  0  &  1.86$\pm$0.12  &  0.342  & $0.8_{-0.4}^{+0.6}$  &  BLL  & E \\
102  &  1RXS J220708.5+360935  &  331.773  &  36.124  &  3(3) &  0  &  1.99$\pm$0.15  &  -1  & $0.7_{-0.3}^{+0.5}$  &  BCU  &  \\
103  &  PMN J2221-5224  &  335.391  &  -52.431  &  2(3) &  1  &  1.79$\pm$0.03  &  -1  & $0.8_{-0.4}^{+0.6}$  &  BLL  &  \\
104  &  {\bf RGB J2247+442 } &  341.974  &  44.226  &  5(6) &  0  &  1.74$\pm$0.06  &  -1  & $1.4_{-0.5}^{+0.7}$  &  BLL  &  \\
105  &  WISEA J230940.84-363248.7  &  347.43  &  -36.547  &  2(3) &  0  &  2.01$\pm$0.06  &  -1  & $0.8_{-0.4}^{+0.6}$  &  BLL  &  \\
106  & {\bf  TXS 2320+343 } &  350.685  &  34.612  &  4(4) &  0  &  2.01$\pm$0.07  &  0.098  & $1.0_{-0.4}^{+0.6}$  &  BLL  &  \\
107  & {\bf  NVSS J232914+375414}  &  352.318  &  37.921  &  3(4) &  0  &  1.88$\pm$0.05  &  0.21  & $1.0_{-0.4}^{+0.6}$  &  BLL  &  \\
108  & {\bf  1RXS J234051.4+801513}  &  355.208  &  80.26  &  2(6) &  0  &  1.91$\pm$0.03  &  0.274  & $1.1_{-0.4}^{+0.6}$  &  BLL  &  \\
109  & {\bf  1RXS J234332.5+343957}  &  355.906  &  34.64  &  3(4) &  0  &  1.69$\pm$0.07  &  0.366  & $1.0_{-0.4}^{+0.6}$  &  BLL  & E \\
110  & {\bf  TXS 2344+068}  &  356.679  &  7.093  &  2(6) &  1  &  1.9$\pm$0.04  &  0.171  & $1.6_{-0.5}^{+0.8}$  &  BLL  &  \\
\hline
\end{longtable}

\begin{longtable}{lllllllllll}
\caption{Known VHE sources detected at $3\sigma$. Notations are the same as in Table \ref{tab:known_sources}.}\\
\label{tab:known_supp}
&4FGL               & RA & Dec & $N_{100}$  &$N_{300}$  & HR   &$z$ & $F_{100}$ &Type & HSP \\ 
\hline
\endfirsthead
\caption{continued.}\\
\hline
&4FGL               & RA & Dec & $N_{100}$  &$N_{300}$  & $\Gamma$   &$z$ & $F_{100}$ &Type & HSP  \\ 
\hline
\endhead
\hline
\endfoot
\hline
1  &  SHBL J001355.9-185406  &  3.48  &  -18.912  &  2(2) &  1  &  1.9$\pm$0.08  &  0.095  & $0.6_{-0.3}^{+0.6}$  &  BLL  & E \\
2  &  RBS 0723  &  131.812  &  11.569  &  2(2) &  0  &  1.77$\pm$0.06  &  0.198  & $0.5_{-0.3}^{+0.5}$  &  BLL  & E \\
3  &  OJ 287  &  133.707  &  20.116  &  2(2) &  0  &  2.21$\pm$0.01  &  0.306  & $0.5_{-0.3}^{+0.5}$  &  BLL  &  \\
4  &  M 82  &  148.947  &  69.667  &  1(3) &  1  &  2.21$\pm$0.03  &  0.001  & $0.6_{-0.3}^{+0.5}$  &  SBG  &  \\
5  &  4C +21.35  &  186.228  &  21.381  &  1(4) &  0  &  2.33$\pm$0.01  &  0.432  & $1.0_{-0.4}^{+0.7}$  &  FSRQ  &  \\
6  &  1ES 1440+122  &  220.698  &  12.013  &  1(4) &  0  &  1.76$\pm$0.05  &  0.163  & $1.0_{-0.4}^{+0.6}$  &  BLL  & E \\
\hline
\end{longtable}

\begin{longtable}{lllllllllll}
\caption{New VHE sources detected at $3\sigma$.}\\
\label{tab:new_supp}
&4FGL               & RA & Dec & $N_{100}$  &$N_{300}$  & HR   &$z$ & $F_{100}$ &Type &HSP  \\ 
\hline
\endfirsthead
\caption{continued.}\\
\hline
&4FGL               & RA & Dec & $N_{100}$  &$N_{300}$  & $\Gamma$   &$z$ & $F_{100}$ &Type & HSP  \\ 
\hline
\endhead
\hline
\endfoot
\hline
1  &  4FGL J0032.3-5539  &  8.097  &  -55.664  &  2(2) &  0  &  1.51$\pm$0.23  &  -1  & $0.5_{-0.3}^{+0.5}$  &  UNID  &  \\
2  &  PMN J0039-2220  &  9.778  &  -22.328  &  1(4) &  1  &  1.78$\pm$0.09  &  0.064  & $1.1_{-0.4}^{+0.7}$  &  BLL  &  \\
3  &  B3 0037+405  &  10.099  &  40.836  &  1(3) &  1  &  1.88$\pm$0.11  &  0.24  & $0.7_{-0.3}^{+0.6}$  &  BLL  & E \\
4  &  GB6 J0045+2127  &  11.34  &  21.467  &  2(2) &  0  &  1.82$\pm$0.03  &  2.07  & $0.5_{-0.3}^{+0.5}$  &  BLL  &  \\
5  &  GB6 J0045+1217  &  11.431  &  12.292  &  1(3) &  0  &  1.95$\pm$0.04  &  -1  & $0.8_{-0.3}^{+0.6}$  &  BLL  &  \\
6  &  4FGL J0054.7-2455  &  13.694  &  -24.933  &  2(2) &  0  &  1.78$\pm$0.04  &  0.61  & $0.5_{-0.3}^{+0.5}$  &  BLL  &  \\
7  &  1ES 0120+340  &  20.791  &  34.354  &  2(2) &  0  &  1.69$\pm$0.08  &  0.272  & $0.5_{-0.2}^{+0.5}$  &  BLL  & E \\
8  &  NVSS J012340+421010  &  20.934  &  42.165  &  2(2) &  0  &  1.89$\pm$0.2  &  -1  & $0.5_{-0.2}^{+0.5}$  &  BLL  & E \\
9  &  SUMSS J014347-584550  &  25.948  &  -58.772  &  2(2) &  0  &  1.83$\pm$0.03  &  -1  & $0.5_{-0.3}^{+0.5}$  &  BLL  &  \\
10  &  1RXS J015658.6-530208  &  29.233  &  -53.031  &  2(2) &  0  &  1.78$\pm$0.05  &  0.25  & $0.5_{-0.3}^{+0.5}$  &  BLL  & E \\
11  &  WN B0210.3+7540  &  33.828  &  75.919  &  2(2) &  2  &  1.8$\pm$0.12  &  0.15  & $0.4_{-0.2}^{+0.4}$  &  BCU  &  \\
12  &  RBS 0334  &  39.424  &  -36.042  &  2(2) &  0  &  1.9$\pm$0.08  &  0.411  & $0.5_{-0.3}^{+0.5}$  &  BLL  &  \\
13  &  FIRST J030013.1-051514  &  45.075  &  -5.247  &  2(2) &  0  &  2.15$\pm$0.25  &  -1  & $0.5_{-0.3}^{+0.5}$  &  BCU  &  \\
14  &  NGC 1218  &  47.111  &  4.118  &  2(2) &  0  &  2.01$\pm$0.05  &  0.03  & $0.5_{-0.3}^{+0.5}$  &  RDG  &  \\
15  &  1RXS J034424.5+343016  &  56.121  &  34.547  &  2(2) &  0  &  1.92$\pm$0.15  &  -1  & $0.5_{-0.3}^{+0.5}$  &  BCU  &  \\
16  &  MS 0419.3+1943  &  65.587  &  19.862  &  1(3) &  0  &  1.92$\pm$0.09  &  0.516  & $0.8_{-0.3}^{+0.6}$  &  BLL  & E \\
17  &  PKS 0422+00  &  66.195  &  0.603  &  2(2) &  0  &  2.2$\pm$0.03  &  0.268  & $0.5_{-0.3}^{+0.5}$  &  BLL  &  \\
18  &  PMN J0427-1835  &  66.886  &  -18.533  &  2(2) &  0  &  1.75$\pm$0.2  &  -1  & $0.5_{-0.3}^{+0.5}$  &  BCU  & E \\
19  &  GB6 J0442+6140  &  70.678  &  61.705  &  2(2) &  0  &  1.96$\pm$0.08  &  0.18  & $0.4_{-0.2}^{+0.4}$  &  BLL  & E \\
20  &  PKS 0454-234  &  74.261  &  -23.415  &  2(2) &  0  &  2.19$\pm$0.01  &  1.0  & $0.5_{-0.3}^{+0.5}$  &  FSRQ  &  \\
21  &  4FGL J0505.6+0415  &  76.404  &  4.266  &  2(2) &  0  &  1.92$\pm$0.07  &  0.424  & $0.5_{-0.3}^{+0.5}$  &  BLL  &  \\
22  &  PKS 0521-36  &  80.737  &  -36.469  &  1(3) &  1  &  2.45$\pm$0.01  &  0.055  & $0.7_{-0.3}^{+0.6}$  &  BCU  &  \\
23  &  1RXS J053629.4-334302  &  84.115  &  -33.723  &  1(3) &  0  &  1.81$\pm$0.04  &  -1  & $0.7_{-0.3}^{+0.6}$  &  BLL  &  \\
24  &  NVSS J064933-313917  &  102.395  &  -31.657  &  2(2) &  0  &  1.71$\pm$0.07  &  -1  & $0.5_{-0.2}^{+0.5}$  &  BLL  &  \\
25  &  1RXS J070132.1+250950  &  105.377  &  25.196  &  2(2) &  0  &  1.91$\pm$0.14  &  -1  & $0.5_{-0.3}^{+0.5}$  &  BCU  & E \\
26  &  GB6 J0706+3744  &  106.63  &  37.738  &  2(2) &  0  &  1.91$\pm$0.03  &  -1  & $0.5_{-0.3}^{+0.5}$  &  BLL  &  \\
27  &  GB6 J0708+2241  &  107.277  &  22.685  &  1(4) &  1  &  2.03$\pm$0.04  &  -1  & $1.1_{-0.4}^{+0.7}$  &  BLL  &  \\
28  &  GB6 J0712+5033  &  108.188  &  50.551  &  2(2) &  0  &  2.21$\pm$0.03  &  0.502  & $0.5_{-0.2}^{+0.5}$  &  BLL  &  \\
29  &  4FGL J0737.4+6535  &  114.374  &  65.591  &  2(2) &  0  &  2.09$\pm$0.16  &  -1  & $0.4_{-0.2}^{+0.4}$  &  UNID  &  \\
30  &  4FGL J0747.5-4927  &  116.882  &  -49.461  &  1(3) &  1  &  2.11$\pm$0.07  &  -1  & $0.7_{-0.3}^{+0.6}$  &  BLL  &  \\
31  &  4C +54.15  &  118.253  &  53.889  &  2(2) &  0  &  2.13$\pm$0.03  &  0.2  & $0.5_{-0.2}^{+0.5}$  &  BLL  &  \\
32  &  4FGL J0806.5+5930  &  121.65  &  59.514  &  2(2) &  0  &  2.04$\pm$0.08  &  0.3  & $0.5_{-0.2}^{+0.5}$  &  BLL  &  \\
33  &  B2 0806+35  &  122.422  &  34.925  &  2(2) &  0  &  1.89$\pm$0.09  &  0.082  & $0.5_{-0.3}^{+0.5}$  &  BLL  &  \\
34  &  PMN J0827-0708  &  126.762  &  -7.144  &  1(3) &  0  &  1.77$\pm$0.07  &  0.247  & $0.8_{-0.4}^{+0.6}$  &  BLL  &  \\
35  &  RX J0830.1+5230  &  127.511  &  52.531  &  2(2) &  0  &  1.78$\pm$0.13  &  0.205  & $0.5_{-0.2}^{+0.5}$  &  BLL  & E \\
36  &  4FGL J0840.1-0225  &  130.039  &  -2.433  &  2(2) &  0  &  1.51$\pm$0.16  &  -1  & $0.5_{-0.3}^{+0.5}$  &  BCU  &  \\
37  &  B3 0850+443  &  133.593  &  44.149  &  2(2) &  0  &  1.9$\pm$0.06  &  0.634  & $0.5_{-0.2}^{+0.5}$  &  BLL  &  \\
38  &  GB6 J0929+5013  &  142.327  &  50.235  &  1(3) &  0  &  2.1$\pm$0.03  &  0.37  & $0.7_{-0.3}^{+0.6}$  &  BLL  &  \\
39  &  1ES 0927+500  &  142.625  &  49.858  &  2(2) &  0  &  1.85$\pm$0.13  &  0.187  & $0.5_{-0.2}^{+0.5}$  &  BLL  & E \\
40  &  RX J0940.3+6148  &  145.121  &  61.816  &  2(2) &  0  &  1.82$\pm$0.08  &  0.211  & $0.4_{-0.2}^{+0.4}$  &  BLL  &  \\
41  &  PMN J0953-0840  &  148.263  &  -8.67  &  2(2) &  0  &  1.89$\pm$0.02  &  0.59  & $0.5_{-0.3}^{+0.5}$  &  BLL  &  \\
42  &  1RXS J095812.8-675241  &  149.534  &  -67.894  &  1(3) &  0  &  2.58$\pm$0.14  &  0.21  & $0.8_{-0.3}^{+0.6}$  &  BCU  &  \\
43  &  B3 1009+427  &  153.194  &  42.482  &  2(2) &  1  &  1.8$\pm$0.06  &  0.365  & $0.5_{-0.2}^{+0.5}$  &  BLL  &  \\
44  &  1RXS J104057.7+134216  &  160.266  &  13.71  &  2(2) &  0  &  1.91$\pm$0.12  &  0.557  & $0.5_{-0.3}^{+0.5}$  &  BLL  &  \\
45  &  PMN J1042-0558  &  160.477  &  -5.953  &  2(2) &  0  &  1.85$\pm$0.11  &  -1  & $0.6_{-0.3}^{+0.6}$  &  BLL  &  \\
46  &  4C +01.28  &  164.624  &  1.564  &  1(3) &  0  &  2.23$\pm$0.01  &  0.185  & $0.8_{-0.4}^{+0.6}$  &  BLL  &  \\
47  &  TXS 1055+567  &  164.665  &  56.463  &  2(2) &  0  &  1.97$\pm$0.01  &  0.143  & $0.5_{-0.2}^{+0.5}$  &  BLL  &  \\
48  &  RX J1100.3+4019  &  165.091  &  40.335  &  2(2) &  0  &  1.85$\pm$0.06  &  0.225  & $0.5_{-0.2}^{+0.5}$  &  BLL  &  \\
49  &  RX J1117.2+0006  &  169.301  &  0.144  &  2(2) &  0  &  1.7$\pm$0.11  &  0.451  & $0.6_{-0.3}^{+0.6}$  &  BLL  & E \\
50  &  RX J1123.8+7230  &  170.959  &  72.513  &  2(2) &  0  &  1.65$\pm$0.11  &  0.38  & $0.4_{-0.2}^{+0.4}$  &  BLL  & E \\
51  &  NVSS J114023+152808  &  175.129  &  15.482  &  2(2) &  0  &  1.73$\pm$0.14  &  0.244  & $0.5_{-0.3}^{+0.5}$  &  BLL  &  \\
52  &  4FGL J1158.8-1430  &  179.709  &  -14.501  &  1(3) &  0  &  1.72$\pm$0.13  &  0.48  & $0.8_{-0.4}^{+0.6}$  &  BCU\footnote{The source is unidentified in 4FGL, but SIMBAD defines it as "blazar candidate"}  &  \\
53  &  4FGL J1203.1+6031  &  180.788  &  60.518  &  2(2) &  0  &  2.12$\pm$0.03  &  0.065  & $0.4_{-0.2}^{+0.4}$  &  BLL  &  \\
54  &  RX J1211.9+2242  &  183.015  &  22.709  &  2(2) &  0  &  1.97$\pm$0.11  &  0.453  & $0.5_{-0.3}^{+0.5}$  &  BLL  & E \\
55  &  PMN J1256-1146  &  194.063  &  -11.776  &  2(2) &  0  &  1.97$\pm$0.05  &  0.058  & $0.6_{-0.3}^{+0.6}$  &  BLL  &  \\
56  &  1RXS J130421.2-435308  &  196.088  &  -43.896  &  1(3) &  0  &  1.89$\pm$0.02  &  -1  & $0.8_{-0.3}^{+0.6}$  &  BLL  &  \\
57  &  1RXS J130737.8-425940  &  196.91  &  -42.995  &  1(3) &  0  &  1.92$\pm$0.03  &  -1  & $0.8_{-0.3}^{+0.6}$  &  BLL  &  \\
58  &  B3 1307+433  &  197.363  &  43.085  &  2(2) &  0  &  1.9$\pm$0.02  &  0.694  & $0.5_{-0.2}^{+0.5}$  &  BLL  &  \\
59  &  TXS 1307-117  &  197.573  &  -11.973  &  1(3) &  0  &  2.01$\pm$0.07  &  -1  & $0.8_{-0.4}^{+0.6}$  &  BLL  &  \\
60  &  RX J1340.4+4410  &  205.147  &  44.161  &  2(2) &  0  &  1.74$\pm$0.1  &  0.546  & $0.5_{-0.2}^{+0.5}$  &  BLL  & E \\
61  &  4FGL J1425.0+3615  &  216.252  &  36.257  &  2(2) &  0  &  1.89$\pm$0.05  &  1.094  & $0.5_{-0.2}^{+0.5}$  &  BLL  &  \\
62  &  B2 1447+27  &  222.396  &  27.769  &  2(2) &  0  &  1.48$\pm$0.11  &  0.031  & $0.5_{-0.2}^{+0.5}$  &  RDG  &  \\
63  &  MS 1458.8+2249  &  225.257  &  22.636  &  2(2) &  0  &  2.02$\pm$0.03  &  0.235  & $0.5_{-0.2}^{+0.5}$  &  BLL  &  \\
64  &  TXS 1516+064  &  229.65  &  6.242  &  1(3) &  0  &  1.8$\pm$0.16  &  0.102  & $0.8_{-0.3}^{+0.6}$  &  RDG  & E \\
65  &  NVSS J152048-034850  &  230.205  &  -3.815  &  2(2) &  0  &  1.78$\pm$0.04  &  -1  & $0.5_{-0.3}^{+0.5}$  &  BLL  &  \\
66  &  4FGL J1547.0-0442  &  236.766  &  -4.708  &  2(2) &  0  &  1.8$\pm$0.16  &  -1  & $0.5_{-0.3}^{+0.5}$  &  UNID  &  \\
67  &  RBS 1558  &  241.592  &  56.498  &  2(2) &  0  &  1.76$\pm$0.1  &  0.45  & $0.4_{-0.2}^{+0.4}$  &  BLL  &  \\
68  &  PKS 1619-765  &  246.655  &  -76.65  &  2(2) &  0  &  1.92$\pm$0.07  &  0.105  & $0.5_{-0.3}^{+0.5}$  &  BLL  &  \\
69  &  NGC 6251  &  247.669  &  82.574  &  2(2) &  0  &  2.35$\pm$0.02  &  0.02  & $0.4_{-0.2}^{+0.4}$  &  RDG  &  \\
70  &  NVSS J165517-224043  &  253.79  &  -22.653  &  1(3) &  0  &  1.89$\pm$0.21  &  -1  & $0.8_{-0.4}^{+0.6}$  &  BCU  &  \\
71  &  NVSS J170433-052839  &  256.138  &  -5.462  &  2(2) &  0  &  1.97$\pm$0.05  &  0.3  & $0.5_{-0.3}^{+0.5}$  &  BLL  &  \\
72  &  1RXS J174420.1+185215  &  266.103  &  18.864  &  2(2) &  0  &  1.88$\pm$0.14  &  0.39  & $0.5_{-0.2}^{+0.5}$  &  BLL  &  \\
73  &  4FGL J1838.4-6023  &  279.601  &  -60.393  &  2(2) &  0  &  2.0$\pm$0.13  &  0.121  & $0.5_{-0.3}^{+0.5}$  &  BLL  &  \\
74  &  TXS 1902+556  &  285.808  &  55.677  &  1(3) &  0  &  1.95$\pm$0.01  &  0.58  & $0.6_{-0.3}^{+0.5}$  &  BLL  &  \\
75  &  PMN J1918-4111  &  289.564  &  -41.189  &  1(3) &  0  &  1.96$\pm$0.02  &  -1  & $0.8_{-0.4}^{+0.6}$  &  BLL  &  \\
76  &  TXS 1922-224  &  291.466  &  -22.341  &  2(2) &  0  &  1.83$\pm$0.12  &  -1  & $0.5_{-0.3}^{+0.5}$  &  BLL  &  \\
77  &  1RXS J195500.6-160328  &  298.777  &  -16.072  &  2(2) &  0  &  2.0$\pm$0.07  &  0.23  & $0.5_{-0.3}^{+0.5}$  &  BLL  &  \\
78  &  4FGL J1955.3-5032  &  298.833  &  -50.54  &  2(2) &  0  &  2.38$\pm$0.12  &  -1  & $0.6_{-0.3}^{+0.6}$  &  UNID  &  \\
79  &  PMN J2014-0047  &  303.599  &  -0.792  &  1(3) &  1  &  2.0$\pm$0.08  &  0.231  & $0.8_{-0.3}^{+0.6}$  &  BLL  &  \\
80  &  RX J2030.8+1935  &  307.741  &  19.599  &  2(2) &  0  &  1.91$\pm$0.04  &  0.27  & $0.5_{-0.2}^{+0.5}$  &  BLL  &  \\
81  &  4FGL J2042.7+1519  &  310.696  &  15.329  &  2(2) &  0  &  2.02$\pm$0.1  &  -1  & $0.5_{-0.2}^{+0.5}$  &  BCU\footnote{The source is classified as "Quasar" in SIMBAD, while it is an unidentified in 4FGL.}  &  \\
82  &  RGB J2054+002  &  313.725  &  0.257  &  2(2) &  0  &  1.9$\pm$0.1  &  0.151  & $0.5_{-0.3}^{+0.5}$  &  BLL  &  \\
83  &  4FGL J2104.4+2116  &  316.123  &  21.283  &  2(2) &  1  &  2.01$\pm$0.08  &  0.36  & $0.5_{-0.2}^{+0.5}$  &  BLL\footnote{The source is classified as BL Lac in SIMBAD, but it is unidentified in 4FGL}  &  \\
84  &  NVSS J213430-213032  &  323.641  &  -21.503  &  2(2) &  0  &  2.0$\pm$0.05  &  -1  & $0.6_{-0.3}^{+0.6}$  &  BLL  &  \\
85  &  4FGL J2159.6-4620  &  329.912  &  -46.334  &  1(3) &  0  &  1.92$\pm$0.09  &  0.4  & $0.8_{-0.4}^{+0.6}$  &  BCU\footnote{The source is classified as blazar candidate in SIMBAD, but it is unidentified in 4FGL} &  \\
86  &  NVSS J220941-045111  &  332.438  &  -4.86  &  2(2) &  0  &  1.92$\pm$0.08  &  0.397  & $0.6_{-0.3}^{+0.6}$  &  BLL  &  \\
87  &  1RXS J221058.3+320327  &  332.724  &  32.053  &  2(2) &  1  &  1.39$\pm$0.18  &  -1  & $0.5_{-0.2}^{+0.5}$  &  BCU  &  \\
88  &  4FGL J2232.6-2023  &  338.172  &  -20.391  &  2(2) &  0  &  1.96$\pm$0.12  &  0.386  & $0.6_{-0.3}^{+0.6}$  &  BLL  & E \\
89  &  RBS 1899  &  342.356  &  -13.001  &  1(3) &  0  &  2.37$\pm$0.15  &  0.607  & $0.9_{-0.4}^{+0.7}$  &  BLL  & E \\
90  &  4FGL J2252.0+4031  &  343.01  &  40.523  &  2(2) &  0  &  2.07$\pm$0.04  &  0.229  & $0.5_{-0.2}^{+0.5}$  &  BLL  &  \\
91  &  RGB J2313+147  &  348.508  &  14.753  &  2(2) &  0  &  1.82$\pm$0.05  &  0.164  & $0.5_{-0.3}^{+0.5}$  &  BLL  & E \\
92  &  4FGL J2317.7+2839  &  349.438  &  28.659  &  2(2) &  0  &  2.03$\pm$0.11  &  -1  & $0.5_{-0.2}^{+0.5}$  &  BLL\footnote{The source has been recently classified as BLL \cite{2024MNRAS.530.4626U}.}  &  \\
93  &  4FGL J2321.0-6308  &  350.258  &  -63.137  &  2(2) &  0  &  1.85$\pm$0.11  &  0.2  & $0.5_{-0.3}^{+0.5}$  &  BLL  &  \\
94  &  4FGL J2336.6+2356  &  354.154  &  23.942  &  2(2) &  0  &  1.95$\pm$0.09  &  -1  & $0.5_{-0.3}^{+0.5}$  &  BCU  &  \\
\hline
\end{longtable}

\end{document}